\documentclass[a4paper, 11pt]{article}

\usepackage{graphicx}
\usepackage{ subfig}
\usepackage{color}
\usepackage{bm}
\usepackage{mathrsfs}
\usepackage{ulem}
\usepackage{multirow}
\usepackage{pdflscape}
\usepackage{rotating}
\usepackage{lscape}
\usepackage{pdflscape}

\def\b{\begin{eqnarray}}
\def\e{\end{eqnarray}}
\def\n{\noindent}
\makeatletter
\newcommand*\bigdot{\mathpalette\bigdot@{.5}}
\newcommand*\bigdot@[2]{\mathbin{\vcenter{\hbox{\scalebox{#2}{$\m@th#1\bullet$}}}}}
\makeatother

\begin{document}

\begin{center}
{\huge \textbf{Van der Waals Universe with Adiabatic \vskip.5cm Matter Creation}}

\vspace {10mm}
\noindent
{\large \bf Rossen I. Ivanov and Emil M. Prodanov} \vskip.5cm
{\it School of Mathematical Sciences, Technological University Dublin, Ireland,}
\vskip.1cm
{\it E-Mails: rossen.ivanov@dit.ie, emil.prodanov@dit.ie} \\
\vskip1cm
\end{center}

\vskip4cm
\begin{abstract}
\n
A FRWL cosmological model with perfect fluid comprising of van der Waals gas and dust has been studied in the context of dynamical analysis of a three-component autonomous non-linear dynamical system for the particle number density $n$, the Hubble parameter $H$, and the temperature $T$. Perfect fluid isentropic particle creation at rate proportional to an integer power $\alpha$ of $H$ has been incorporated. The existence of a global first integral allows the determination of the temperature evolution law and hence the reduction of the dynamical system to a two-component one. Special attention is paid to the cases of $\alpha = 2$ and $\alpha = 4$ and these are illustrated with numerical examples. The global dynamics is comprehensively studied for different choices of the values of the physical parameters of the model. Trajectories in the $(n, H)$ phase space are identified for which temporary inflationary regime exists.

\end{abstract}
\vskip1cm
\noindent
{\bf Keywords:} Dynamical systems, FRWL cosmology, accelerated expansion, real gas.

\newpage

\section{Introduction}

The acceleration of the cosmic expansion and observational data (Supernov\ae $\,$ Type Ia, Cosmic Microwave Background, Baryon Acoustic Oscillations) are fit best by the current concordance model --- the $\Lambda$CDM model which incorporates Dark Energy, modelled by the cosmological constant $\Lambda$, and cold (pressureless) Dark Matter. There are open issues in relation to such model --- the so called Cosmological Coincidence Problem: it is known observationally that the present values of the densities of dark energy and dark matter are of the same order of magnitude while, under the $\Lambda$CDM model, the dark-energy density is constant and the dark-matter density is proportional to the inverse third power of the scale factor with the ratio of the two densities varying in time from infinity to zero. There are numerous alternative models, not without open issues on their own,  which accommodate acceleration of the cosmic expansion: modified gravity theories, inhomogeneous cosmologies, gravitationally induced particle creation models. In the literature, special attention has been gathered by the adiabatic, or isentropic, production \cite{engoliam, prigo, equiva1, equiva2} of perfect fluid particles in which the specific entropy (entropy per particle) is conserved (with ``isentropic" referring to this). There is overall entropy production due to the enlargement of the phase space of the system as the particle number increases. The imposed condition of conserved specific entropy during the production of perfect fluid particles leads to a simple relationship between the particle production rate and particle ``creation" pressure. Zimdahl \cite{zimdahl} studies cosmological particle production with  production rate which depends quadratically on the Hubble rate $H$ and confirms the existence of solutions which describe a smooth transition from inflationary to non-inflationary behavior. The present work falls in this category and offers a full dynamical analysis of isentropic perfect fluid particle production rate that depends on $H^{\alpha}$ with $\alpha$ being a positive integer. Special attention is paid to the cases of $\alpha = 2$ and $\alpha = 4$, but the analysis can be easily extended to any other integer positive values of $\alpha$, including odd values --- due to the second law of thermodynamics, these work in the regime of expansion only \cite{nie2}. The setting of the proposed model is a flat FRWL Universe with perfect fluid comprising of two fractions: real gas wit van der Waals equation of state and dust and the tools used are those of dynamical systems, see for example, \cite{vilasi,ross}, and as those used in the study of $n$--$H$--$T$ (where $n$ is the particle number density, and $T$ is the temperature) dynamical analysis of cosmological quintessence real gas model with a general equation of state \cite{nie}. The dynamical variables are again $n$, $H$, and $T$, but due to the existence of a global first integral (in addition to second integrals), the temperature evolution law has been easily determined and the dynamical system reduced to a two component one over the $(n, H)$ phase space. Inflationary regime with exit from the inflationary behaviour has been identified, both for $\alpha = 2$ and for $\alpha = 4$, and full classification of the possible phase-space trajectories, subject to the variation of the several physical parameters of the model, has been provided.

\section{The Model}

This paper studies a Universe modelled classically as a fluid comprising of a binary mixture of dust with energy density $\rho_d$ and pressure
$p_d = 0$ and a van der Waals gas with equation of state
\b
\label{eos}
p = n T [1 + n F(T)],
\e
where $p$ is the pressure, $T$ is the temperature, $n = N / V$ --- the number of particles $N$ per unit volume $V$ --- is the particle number density
and $F(T)$ is the term describing two-particle interaction: $F(T) = A - B/T$, where $A$ and $B$ are positive constants\footnote{To aid the analysis, a
numerical example is presented in this work. It is for van der Waals gas, the parameters of which are $A = 1/100$ and $B = 10$.}. \\
The Universe is described, using Planck units, by the $\!$ flat Friedmann--Robertson--Walker--Lema\^itre metric:
\b
ds^2 = g_{\mu \nu} dx^\mu dx^\nu = dt^2 - a^2(t) [dr^2 + r^2 (d \theta^2 + \sin^2 \theta \, d \phi^2)],
\e
where $a(t)$ is the scale factor of the Universe. \\
The particle number is not conserved due to a process of particle creation and annihilation \cite{engoliam, prigo}. This process manifests itself,
geometro-thermodynamically \cite{equiva1, equiva2}, through the appearance of ``creation pressure'' $\Pi$ in the cumulative energy-momentum tensor $T_{\mu \nu}$
\cite{zimdahl,viscosity, zimdahl2}:
\b
\label{te}
T_{\mu \nu} = (\rho + \rho_d + p + \Pi) \, u_\mu \, u_\nu - (p + \Pi) \, g_{\mu \nu} \, ,
\e
where $u^\mu = dx^\mu / d \tau$ (with $\tau$ being the proper time) is the flow vector satisfying $g_{\mu \nu} u^\mu u^\nu \!= 1$. \\
The Friedmann equations are:
\b
\label{h1}
\frac{\ddot{a}}{a} & = & - \frac{1}{6} [ \rho_d + \rho + 3 (p + \Pi) ], \\
\label{h2}
H^2 & = & \frac{1}{3} (\rho_d + \rho),
\e
where $H(t) = \dot{a}(t) / a(t)$, the Hubble parameter, will be considered as one of three dynamical variables of a three-component autonomous
dynamical system, also involving the particle number density $n(t)$ and the temperature $T(t)$. \\
Combining (\ref{h1}) and (\ref{h2}), yields:
\b
\label{dynami}
\dot{H} =  - \frac{3}{2} H^2 - \frac{1}{2} (p + \Pi).
\e
The continuity equation for the particles of the perfect fluid is $N^\mu_{\phantom{\mu }; \mu} = n \Gamma,$ where $N^\mu = n u^\mu$ is the particle
flow vector and $\Gamma$, the particle production rate, is an input quantity in the phenomenological description \cite{zimdahl}. In this work, the
dynamics of a model with particle production rate \cite{freaza}:
\b
\label{Psi}
\Gamma = 3 \beta H^\alpha,
\e
where $\beta$ is a constant, will be studied. As will be shown shortly, due to the second law of thermodynamics, one must have $\Gamma > 0$ so that
the entropy is never decreasing. \\
With such particle production rate, the particle conservation equation reads off as
\b
\label{nn}
\dot{n} = - 3 n H + n \Gamma = - 3 n H (1 - \beta H^{\alpha -1}).
\e
This equation will be further used as one of the evolution equations of the dynamical system. \\
The energy conservation equation for the van der Waals gas and for the dust are
\b
\label{h3}
\dot{\rho} + 3H(\rho + p + \Pi) & = & 0, \\
\label{ce-d}
\dot{\rho_d} + 3H \rho_d  & = & 0,
\e
respectively. \\
The separate conservation laws stipulate that there would be no exchange between the two components of the Universe. \\
The ``creation pressure" $\Pi$, in the case of conserved specific entropy $s$ (i.e. entropy per particle, $s = S/N$, where $S$ is the total entropy),
is given by \cite{prigo}:
\b
\label{Pi}
\Pi = -\frac{\Gamma (\rho + p)}{3H} = - \beta \, (\rho + p) \, H^{\alpha - 1}.
\e
Note that the total entropy $S$ is not conserved due to the enlargement of the phase space resulting from the particle production \cite{prigo}. \\
On the issue of the equivalence of bulk viscosity and matter creation, Calva\~o {\it et al.} \cite{equiva1} and Lima {\it et al.} \cite{equiva2} argue that the matter creation process, as described by Prigogine \cite{prigo}, can generate the same dynamic behavior as a FRWL universe with bulk viscosity, while the models being quite different from a thermodynamic point of view. Brevik {\it et al.} \cite{br3} conclude that creation and viscosity concepts do not describe one and the same physical process --- it is shown that viscous and creation universes can develop dynamically in the same manner, but the thermodynamic requirement for their identification is violated. The dynamic pressure $\Pi$ in case of bulk viscosity is given by $\Pi = - 3 \zeta H$, where $\zeta$ is the bulk viscosity co-efficient \cite{equiva1, equiva2, br3}, while in the case of matter creation processes, similar arguments lead to $\Pi = - \alpha n \Gamma / (3H)$, where $\alpha$ is a phenomenological co-efficient, called creation co-efficient, and it is closely related to the creation process --- see \cite{equiva1, equiva2, br3} and the references therein. The adiabaticity of the fluid, namely, the conservation of the specific entropy, $\dot{s} = 0$, leads to the dependence on time of the creation co-efficient $\alpha$: one gets $\alpha = (\rho + p)/n$ --- see \cite{br3} --- and with this, $\Pi = - \alpha n \Gamma / (3H)$ becomes the same as (\ref{Pi}). \\
Substituting (\ref{Pi}) into (\ref{dynami}) gives:
\b
\label{hashdot}
\dot{H} =  - \frac{3}{2} H^2 - \frac{1}{2} \left[ p(n,T) (1 - \beta H^{\alpha - 1}) - \beta \rho(n, T) H^{\alpha -1} \right].
\e
This equation describes the dynamical evolution of the Hubble parameter and will be the second equation of the dynamical system. \\
The dynamical equation (\ref{nn}), multiplied across by $a^3$, reads off as $dN/dt = (d/dt) (a^3 n) = a^3 n \Gamma$.
Differentiating separately $N = n a^3$ with respect to time, using $\dot{a} = a H$ and (\ref{nn}) gives $\dot{N} = 3 \beta N H^\alpha$.
Also, from $s = S/N = $ const, one gets $\dot{S} / S = \dot{N} / N = 3 \beta H^\alpha$.  Thus, the constant $\beta$ will be taken as positive and
$\alpha$ will be taken as a positive even integer. In the analysis, $\alpha$ and $\beta$ will be considered as parameters of the model. \\
The integrability condition for the Gibbs equation
\b
T ds = d \Bigl({\rho \over n}\Bigr) + p \, d \Bigl({1\over n}\Bigr) =
-\left({\rho + p \over n^2}\right) \,\, dn + {1 \over n} \,\, d\rho.
\e
is
\b
\label{integrability}
n \biggl(\frac{\partial T}{\partial n}\biggr)_\rho + (\rho + p)\biggl(\frac{\partial T}{\partial \rho}\biggr)_n
= T \biggl(\frac{\partial p}{\partial \rho}\biggr)_n.
\e
The latter can be written as the following thermodynamic identity:
\b
\label{tdi}
\rho + p = T \biggl( \frac{\partial p}{\partial T}\biggr)_n + n \biggl( \frac{\partial \rho}{\partial n}\biggr)_T.
\e
In thermodynamical variables $n$ and $T$, the time evolution of the energy density is:
\b
\label{tua}
\dot{\rho}(n, T) = \biggl(\frac{\partial \rho}{\partial n}\biggr)_T \dot{n} + \biggl(\frac{\partial \rho}{\partial T}\biggr)_n \,\, \dot{T}.
\e
On the other hand, the energy conservation equation for the van der Waals gas can be written as:
\b
\label{h33}
\dot{\rho}(n, T) = (\rho + p) (\Gamma - 3H) = - 3 (\rho + p) H(1 - \beta H^{\alpha-1}).
\e
Using the number conservation equation (\ref{nn}) in (\ref{tua}) and equating to (\ref{h33}) gives:
\b
- 3 (\rho + p) H(1 - \beta H^{\alpha-1})
 = -3 n H(1 - \beta H^{\alpha-1}) \, \biggl(\frac{\partial \rho}{\partial n}\biggr)_T
+ \biggl(\frac{\partial \rho}{\partial T}\biggr)_n \,\, \dot{T}.
\e
Expressing $\rho + p$ from (\ref{tdi}) and substituting in the above gives the temperature evolution law:
\b
\label{T}
\dot{T} = - 3 H (1 - \beta H^{\alpha-1}) T \biggl( \frac{\partial p}{\partial \rho} \biggr)_n
= - 3 H (1 - \beta H^{\alpha-1}) T \frac{\left( \frac{\partial p}{\partial T} \right)_n}{\Bigl( \frac{\partial \rho}{\partial T} \Bigr)_n}
\e
and third dynamical equation of the system. \\
In the absence of particle creation or annihilation (i.e. when $\beta = 0$), the above reduces to the well known form given in \cite{nie, maa, lima}.\\
Using the equation of state (\ref{eos}) for the van der Waals gas,
\b
p(n,T) = n T (1 + An) - B n^2,
\e
one finds $(\partial p / \partial T)_n
= n(1 + An).$
Substituting this and the equation of state into the thermodynamic identity (\ref{tdi}) yields:
\b
\biggr[ \frac{\partial}{\partial n} \biggl( \frac{\rho}{n} \biggr) \biggr]_T = - B.
\e
This differential equation can be easily integrated:
\b
\label{ejgo}
\rho = n [ \phi(T) - B n].
\e
In the case of an ideal monoatomic gas with three translational degrees of freedom, the mass density is, approximately, $\rho = n [m_0 + (3/2) T]$.
The expression (\ref{ejgo}) for $\rho$ should agree with that for an ideal gas when ideal gas limit is applied for the van der Waals gas, that is,
when $A$ and $B$ are both set to zero. This gives $\phi(T) = m_0 + (3/2) T$. Namely, the energy density $\rho$, the number density $n$, and the
temperature $T$ of the van der Waals gas are related via
\b
\label{rvg}
\rho(n, T) = n (m_0 + \frac{3}{2} T) - B n^2.
\e
Thus, $(\partial \rho / \partial T)_n = (3/2) n$ and the dynamical system for the case of a van der Waals gas becomes:
\b
\label{1}
\!\!\dot{n} \!\!\! & = & \!\!\!\! - 3 n H (1 - \beta H^{\alpha - 1}), \\
\label{2}
\!\!\dot{H} \!\!\! & = & \!\!\!\! - \frac{3}{2} H^2 - \frac{1}{2} \left[ (1 - \beta H^{\alpha - 1}) p(n, T)
- \beta  H^{\alpha - 1} \rho(n, T) \right],  \\
\label{3}
\!\!\dot{T} \!\!\! & = & \!\!\!\! - 2 (1 + A n) H (1 - \beta H^{\alpha - 1}) T,
\e
where $p(n, T) =  n T ( 1 + A n ) - B n^2$ and $\rho(n, T) =  n [m_0 + (3/2) T] - B n^2$. \\
There is a symmetry: dividing (\ref{3}) by (\ref{1}) gives:
\b
\label{asb}
\frac{dT}{dn} = \frac{2T(1+An)}{3n} > 0 \, \mbox{ as $n > 0$,}
\e
and this is independent of $H$. \\
Equation (\ref{asb}) can be easily integrated to get the temperature evolution law in terms of the particle number density:
\b
\label{t}
T(n) = \tau \, n^{\frac{2}{3}} \, e^{\frac{2 A n}{3}},
\e
where the positive constant $\tau$ represent a temperature scale and is a third parameter of the model (in addition to $\alpha$ and $\beta$). Note that the temperature is independent of $\alpha$ and $\beta$. \\
Equation (\ref{asb}) and its solution are the same as the ones encountered in the case of absence of matter creation or annihilation \cite{nie}. \\
The temperature can be excluded so that the system can be reduced to a two-component one:
\b
\label{2d-1}
\dot{n} & \equiv f_1(n, H) & = 3 n H (\beta H^{\alpha -1} - 1),  \\
\label{2d-2}
\dot{H}  & \equiv f_2(n, H) & = \, - \frac{3}{2} \, H^2 + \frac{1}{2} \, \tau \, n^{\frac{5}{3}} \, e^{\frac{2An}{3}}
\left[ (\beta H^{\alpha-1} - 1) \left(An + \frac{5}{2}\right) + \frac{3}{2} \right] \nonumber \\
&& \hskip3.15cm + \, \frac{1}{2} \, \beta \, (m_0 - 2Bn)\, n \, H^{\alpha-1} + \frac{1}{2} \, B \, n^2.
\e

\section{Analysis}

There is a {\it global} first integral given by:
\b
\label{i}
I(n,T) = T \, n^{-\frac{2}{3}} \, e^{-\frac{2 A n}{3}} = \tau \mbox{ = const } > 0.
\e
A {\it second integral} $K(\vec{x}) = 0$ of an autonomous dynamical system of the type $\dot{\vec{x}}(t) = \vec{F}[\vec{x}(t)]$ is defined by $(d/dt)
K(\vec{x}) = \mu(\vec{x}) K(\vec{x})$. It is as an invariant, but only on a restricted subset, given by its zero level set \cite{gor}.  As no
trajectory can cross a hyper-surface defined by a second integral, the second integrals ``fragment" the phase space into regions with separate
dynamics (yet governed by the same dynamical system). For the two-component dynamical system, the ordinate $n = 0$ is one such invariant manif
old because $(d/dt) n = [3H(\beta H^{\alpha-1}-1)] n$. Similarly, the curve defined by $3H^2 - \rho = 3H^2 - n[m_0 + (3/2)T] + Bn^2 = 0$ is another second
integral because $(d/dt)(3H^2 - \rho) = -3H(3H^2 - \rho).$ It will be called a {\it separatrix} --- see Figure 1. \\
There is a value $\tau_0$ of $\tau_0$ for which the separatrix $3H^2 - n[m_0 + (3/2)T] + Bn^2 = 3H^2 - n[m_0 + (3/2) \, \tau \, n^{2/3} \, e^{2An/3}]
+ Bn^2 = 0$ is tangent to the $n$-axis at point, say $n_0$ (see Figure 1). Both $\tau_0$ and $n_0$ can be determined as follows. When $\tau = \tau_0$,
the separatrix has a minimum at $n_0$ and that minimum is $0$. Thus, $(3/2) \, \tau_0 \, n_0^{2/3} \, e^{2An_0/3} = Bn_0 - m$ and $(d/dn) \left[ n
[m_0 + (3/2) \, \tau \, n^{2/3} \, e^{2An/3}] - Bn^2\right]_{n=n_0, \tau = \tau_0} = 0$ with solutions $n_0 = [ 2 m_0 A + B   + (4 m_0^2 A^2 + 20 m_0
A B  + B^2)^{1/2} \, ] / (4AB)$ and $\tau_0 = (2/3) (Bn_0 - m_0) n_0^{-2/3} \, e^{-2An_0/3}$. \\
\begin{figure}[!ht]
\centering
\includegraphics[height=6cm,width=0.5\textwidth]{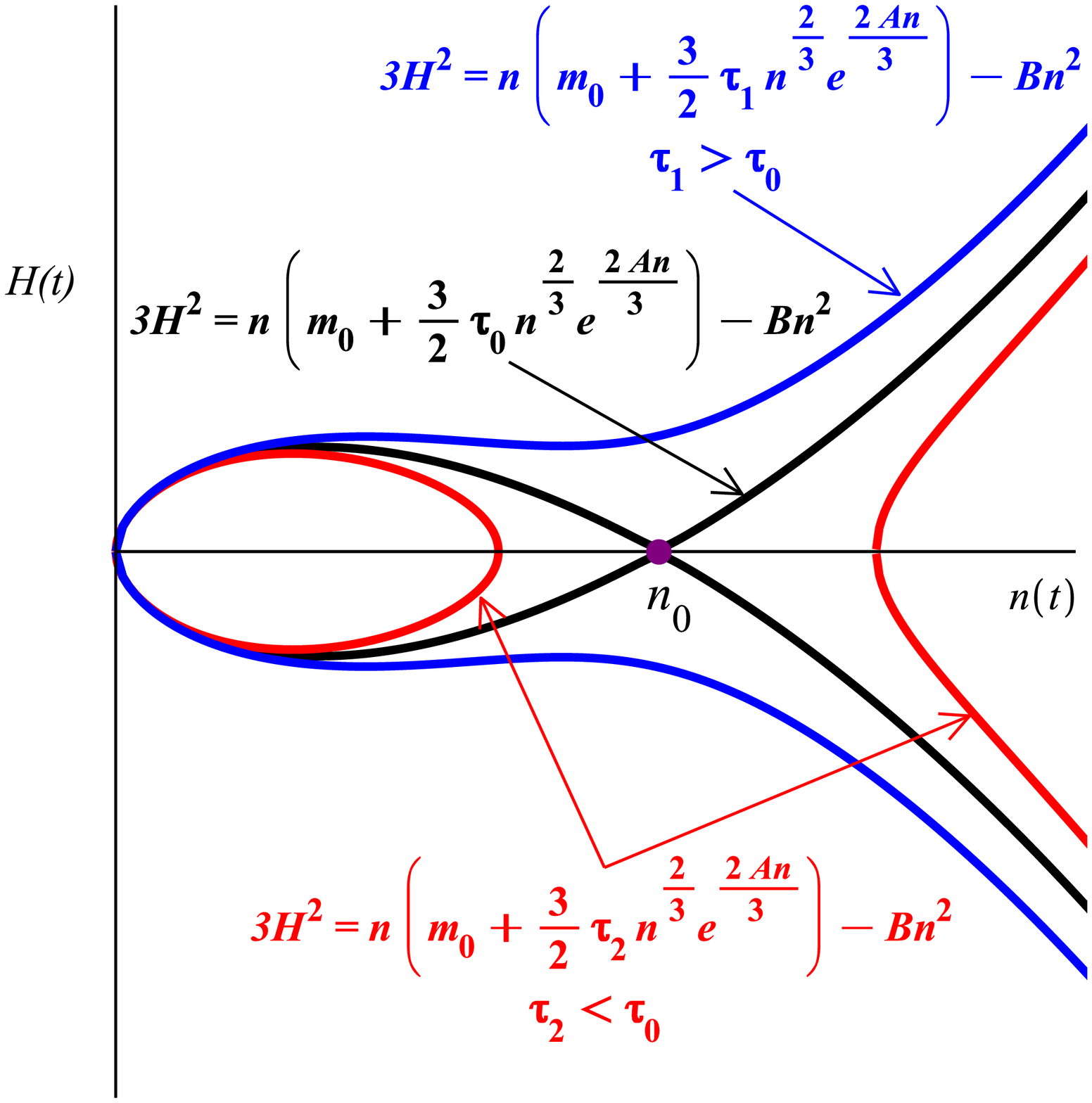}
\caption{\footnotesize{The separatrix $3H^2 - n[m_0 + (3/2) \, \tau \, n^{2/3} \, e^{2An/3}] + Bn^2 = 0$ is an open curve when $\tau > \tau_0 = 14.78$
for a van der Waals gas with parameters $A = 0.01$ and $B = 10$ and $m_0$, the typical mass of a representative particle, taken as $100$. When $\tau <
\tau_0$, the separatrix has a loop at low $n$ and an open part at high $n$.  When $\tau > \tau_0$, the trajectories to the right of the separatrix are
those for dust component $\rho_d < 0$, while those above or below it are with $\rho_d > 0$. On the separatrix itself, $\rho_d = 0$. When $\tau <
\tau_0$, the trajectories to the right of the open curve and those inside the loop are with $\rho_d <0$ while all others have $\rho_d > 0$. The curve
with $\tau = \tau_0$ is tangent to the abscissa at $n_0 = \bigl( 2 m_0 A + B   + \sqrt{4 m_0^2 A^2 + 20 m_0 A B  + B^2} \, \bigr) / (4AB) = 73.59$.
The energy density $\rho[n, T(n)] = n[m_0 + (3/2) \, \tau \, n^{2/3} \, e^{2An/3}] - Bn^2$ is positive for all values of $n$ if $\tau > \tau_0$. }}
\label{F1}
\end{figure}
\n
$\!\!\!\!\!$The energy density $\rho[n, T(n)] = n[m_0 + (3/2) \, \tau \, n^{2/3} \, e^{2An/3}] - Bn^2 > 0$ may become negative over a certain range of $n$,
depending on the choice of initial conditions, namely, depending on $\tau$. Such trajectories would temporarily violate the weak energy condition and,
as this is admissible in phantom cosmology models \cite{car}, the validity of the model will not be restricted by this. \\
The stability matrix $L$ for the two-component dynamical system (\ref{2d-1})--(\ref{2d-2}) is given by:
\b
L_{11} & = & \frac{\partial f_1}{\partial n} = 3 H (\beta H^{\alpha - 1} - 1), \\
L_{12} & = & \frac{\partial f_1}{\partial H} = 3 n (\alpha \beta H^{\alpha - 1} - 1), \\
L_{21} & = & \frac{\partial f_2}{\partial n} = \frac{1}{3} \, \tau \, n^{\frac{2}{3}} \, e^{\frac{2An}{3}} \,
\left[ \left( An + \frac{5}{2} \right)^2 (\beta H^{\alpha - 1} - 1)
+ \frac{3}{2} \, \beta A n H^{\alpha - 1} + \frac{15}{4} \right] \nonumber \\
&& \hskip1.20cm + \,\, \frac{1}{2} \, \beta m_0 H^{\alpha -1} + (1 - 2 \beta H^{\alpha -1})Bn, \\
L_{22} & = & \frac{\partial f_2}{\partial H} = -3 H + \frac{1}{2} \, \beta \, (\alpha - 1) \, \biggl[ \left( An + \frac{5}{2} \right) \, \tau \,
n^{\frac{5}{3}} \, e^{\frac{2An}{3}} +  (m_0 - 2Bn)\,n \biggr] H^{\alpha-2}. \nonumber \\
\e
There are three types of critical points for the dynamical system. Firstly, one has the critical points $(n^*,H^* = 0)$, where $n^*$ are the solutions
of the equation $p(n^*) = 0$, that is $\tau (An^* + 1) n^{*^{5/3}} e^{2An^*/3} - Bn^{*^2} = 0$. This can be written as:
\b
T(n^*) = T^*(n^*)
\e
with
\b
T^*(n^*) = \frac{B n^*}{A n^* + 1}.
\e
Depending on the parameter $\tau$ (i.e. on the choice of initial conditions), the number of intersection points of these two curves is one (the
origin), two [the origin and a point $\widetilde{n^*}$ at which $T(n^*)$ is tangent to $T^*(n^*)$], or three --- one of which is the origin and the
other two are $\nu_{1,2}^*$ which tend to $\widetilde{n^*}$ as $\tau \to \widetilde{\tau}$ from below --- see Figure 2a.\\
To determine the value of $\widetilde{\tau},\,$ for which $T(n^*) = \widetilde{\tau} n^{*^{2/3}} e^{2An^*/3}$ is tangent to $T^*(n^*) =
Bn^*/(An^* + 1)$, and to also determine the point $\widetilde{n^*}$ from the $n^*$-axis where these two curves are tangent to each other, consider the
following. At point $\widetilde{n^*}$, the two curves intersect, i.e. $\widetilde{\tau} \, \widetilde{n}^{*^{2/3}} e^{2A\widetilde{n}^*/3} =
B\widetilde{n}^*/(A\widetilde{n}* + 1)$, and, also, the tangents to the two curves coincide, i.e. $[(d/dn^*) T(n^*)]_{(n^* = \widetilde{n}^*, \tau =
\widetilde{\tau})} = [(d/dn^*) T^*(n^*)]_{(n^* = \widetilde{n}^*, \tau = \widetilde{\tau})} $. From these two simultaneous equations, one
immediately determines that $\widetilde{n}^* = (\sqrt{3/2}-1)/A$ and that $\widetilde{\tau} = B \, \widetilde{n}^{*^{1/3}} e^{-2A\widetilde{n}^*/3} \,
(1+A\widetilde{n}^*)^{-1} = \sqrt{2/3} (\sqrt{3/2}-1)^{1/3} e^{2/3-\sqrt{2/3}} A^{-1/3} B$. (For the numerical example considered, one has
$\widetilde{n}^* = 22.47$ and $\widetilde{\tau} = 19.84$.) \\
\begin{figure}[!ht]
\centering
\subfloat[\scriptsize  ]
{\label{F2a}\includegraphics[height=4.1cm, width=0.32\textwidth]{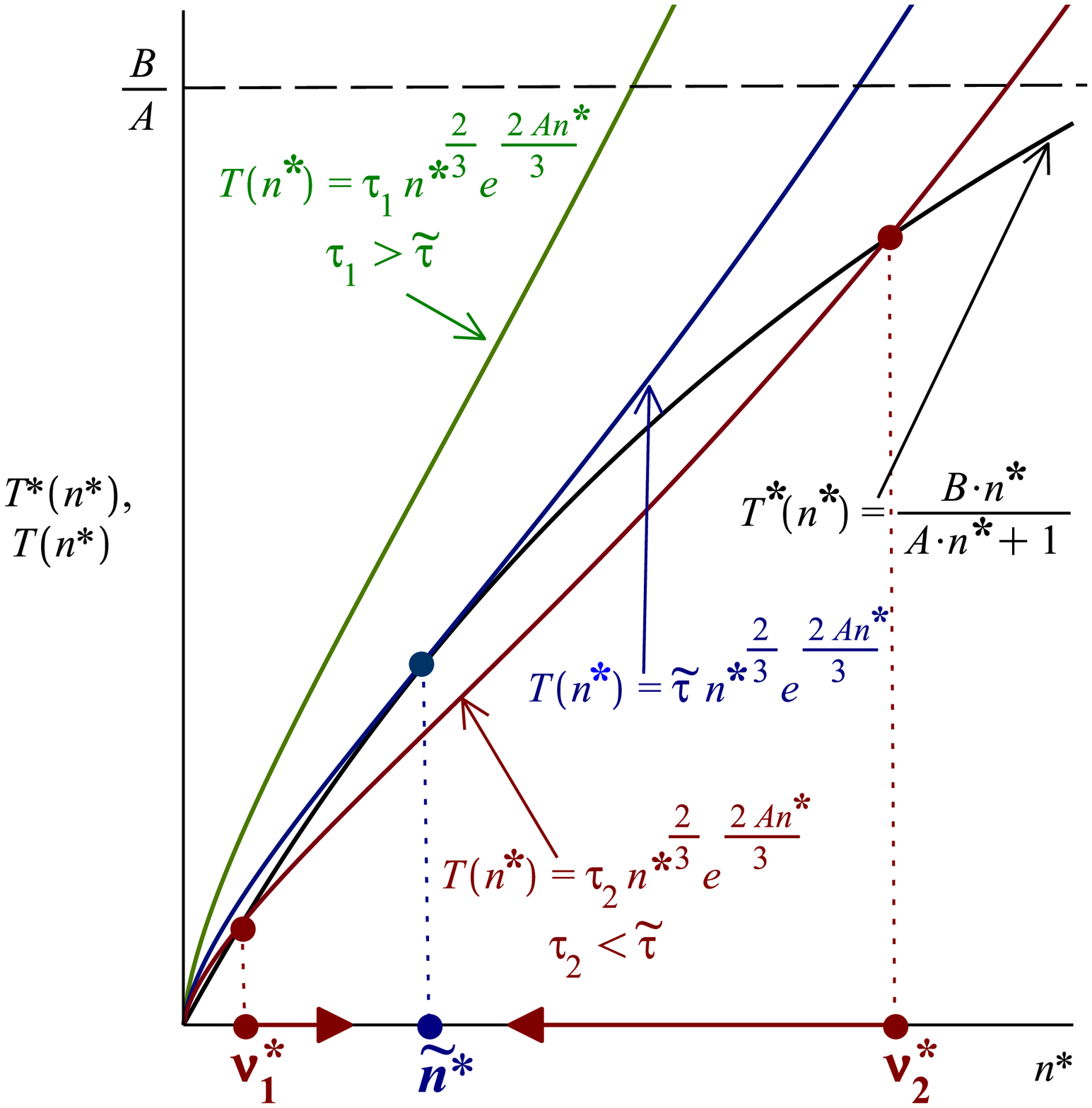}}
\,
\subfloat[\scriptsize ]
{\label{F2b}\includegraphics[height=4.1cm,width=0.32\textwidth]{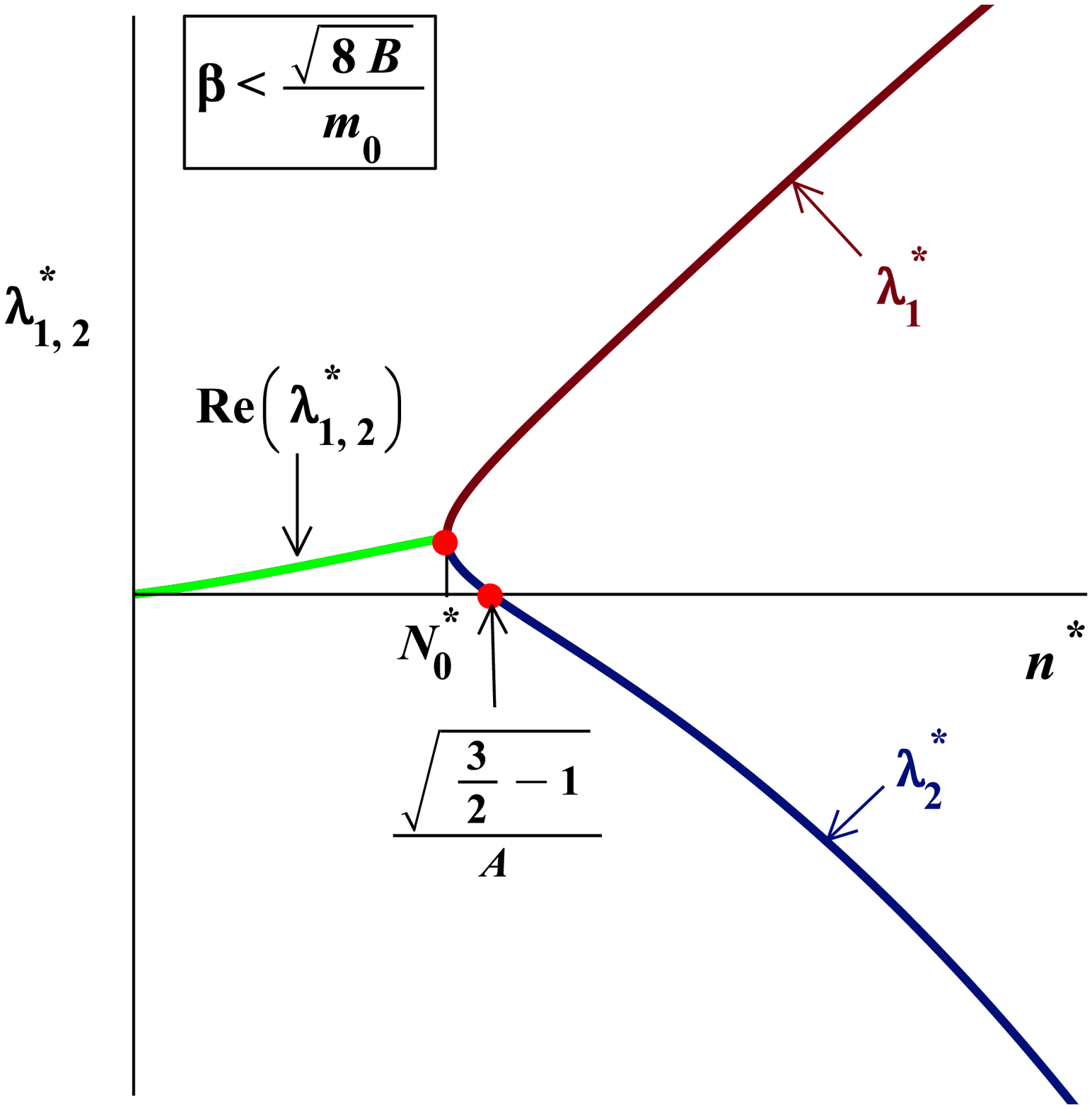}}
\,
\subfloat[\scriptsize ]
{\label{F2c}\includegraphics[height=4.1cm,width=0.32\textwidth]{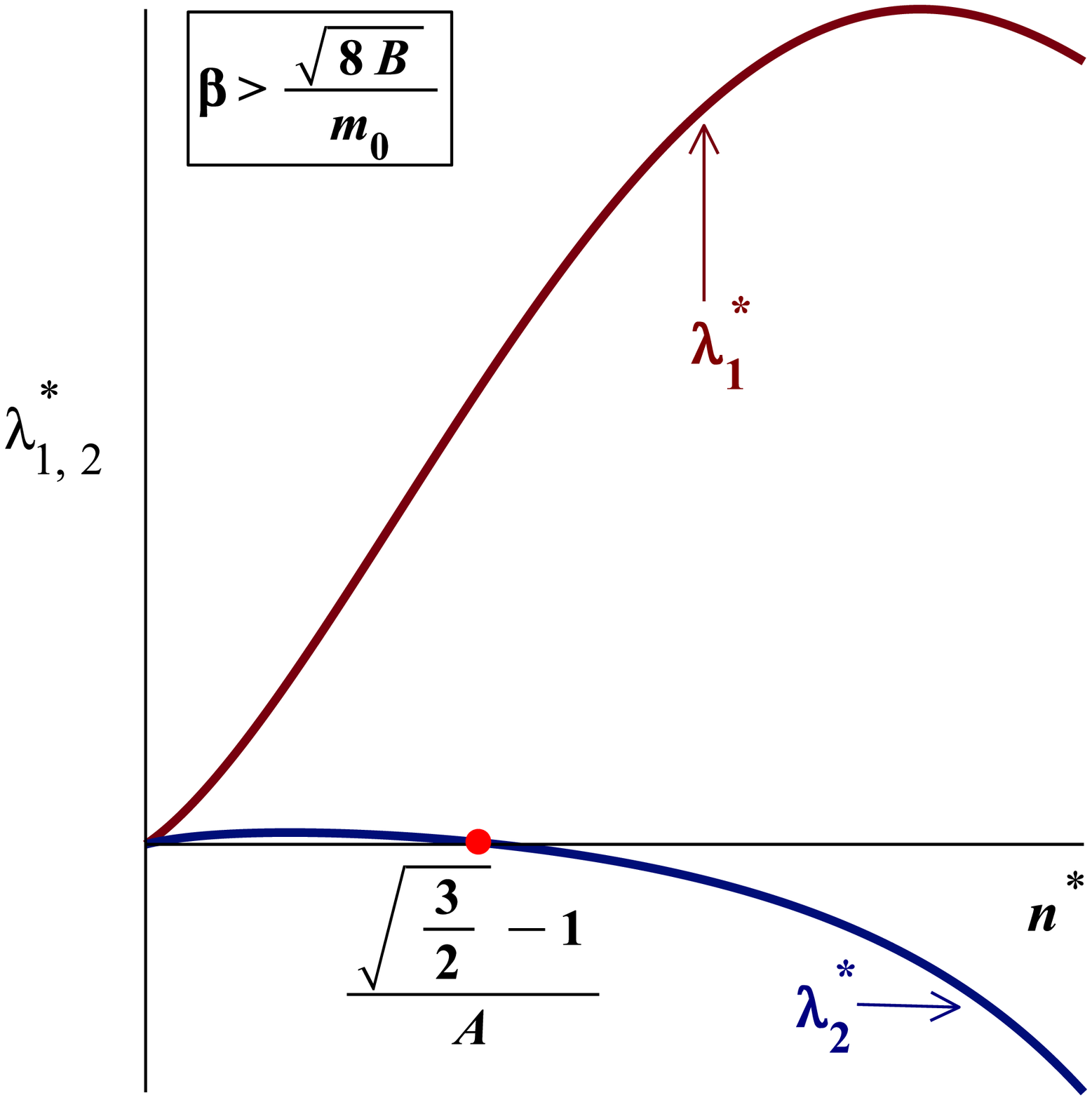}}
\caption{\footnotesize{The critical points of the type $(n^*, H^* = 0)$ have $n^*$ determined by the intersection points of the curve
$T(n^*) = \tau n^{*^{2/3}} e^{2An^*/3}$ with the curve $T^*(n^*) = B n^*/(1 + A n^*)$}. These are: only the origin,
when $\tau > \widetilde{\tau} = \sqrt{2/3} (\sqrt{3/2}-1)^{1/3} e^{2/3-\sqrt{2/3}} A^{-1/3} B$; the origin and $\widetilde{n}^* = (\sqrt{3/2} -
1)/A$ when $\tau = \widetilde{\tau}$; and the origin and $\nu_{1,2}^*$ when $\tau < \widetilde{\tau}$ (with $\nu_{1,2}^* \to \widetilde{n}^*$ when
$\tau \to \widetilde{\tau}$ from below)--- Figure 2a. \\ The eigenvalues $\lambda_{1,2}^*$ at critical points to the left of $\widetilde{n}^* =
(\sqrt{3/2} - 1)/A$ are both positive or with positive real parts (depending on $\beta$), while at critical points to the right of $\widetilde{n}^*$
the eigenvalues are real with $\lambda_1^*$ being positive, while $\lambda_2^*$ --- negative (see Figures 2b and 2c). Given that one of the
eigenvalues is always positive or has positive real part, critical points $(n^*, H^* = 0)$ are never stable.}
\label{Figure2}
\end{figure}
\n
$\!\!\!\!\!$Focusing firstly on the case of $\alpha = 2$, the components of the stability matrix at the critical point $(n^*, H^* = 0)$ are: $L^*_{11} = 0, \,\,\,
L^*_{12} = - 3n^*$,
\b
L_{21}^* & = & - \frac{1}{3} \, \frac{Bn^*}{An^* + 1} \, \left( A^2 n^{*^2} + 2 A n^* - \frac{1}{2} \right), \mbox{ and}\\
L_{22}^* & = & \frac{1}{2} \, \beta \, \frac{n^*}{An^* + 1} \, \left[ - A B n^{*^2} + \left( m_0 A + \frac{B}{2} \right) n^* + m_0 \right].
\e
The eigenvalues at this point are:
\b
\label{www}
\lambda_{1,2}^* = \frac{1}{2} L_{22}^* \,\, \pm \,\, \frac{1}{2} \sqrt{L_{22}^{*^2} - 12 L_{21}^* n^*}.
\e
Note that the point at which $L_{21}^*$ becomes zero, that is, the point at which the smaller eigenvalue $\lambda^*_2$ changes sign, is exactly equal
to the determined earlier $\widetilde{n}^* = (\sqrt{3/2}-1)/A$ --- the point at which $T(n^*)$ is tangent to $T^*(n^*)$ when $\tau =
\widetilde{\tau}$. With the decrease of $\tau$ in $T(n^*)$, the point at which the graphs of $T(n^*)$ and $T^*(n^*)$ are tangent bifurcates into two
intersection points: $\nu_{1,2}^*$ (see Figure 2a). Thus, for critical points to the left of $\widetilde{n}^*$, where $L_{21}^*$ is positive, the
eigenvalues are both positive or with positive real parts, while for critical points to the right of $\widetilde{n}^*$, where $L_{21}^*$ is negative,
the eigenvalues are real with $\lambda_1^*$ being positive, while $\lambda_2^*$ --- negative (see Figures 2b and 2c). In view of this, given that the
eigenvalue $\lambda_1^*$ is always positive over the range of $n^*$ where it is real or it always has positive real part over the range of $n^*$ where
it is complex, critical points $(n^*, H^* = 0)$ are never stable. \\
The eigenvalues $\lambda_{1,2}^*$ will be real numbers when the determinant $L_{22}^{*^2} - 12 L_{21}^* n^*$ is non-negative. This happens when $\beta
>  \sqrt{8B} / m_0 = 0.09$. When $\beta < \sqrt{8B} / m_0$, the eigenvalues will be complex numbers when $n^*$ is in the interval $0 < n^* <
N_0^*$, where $N_0^*$ (which is smaller than $\widetilde{n}^*$) is the only positive root of $L_{22}^{*^2} - 12 L_{21}^* n^* = 0$:
\b
&& \hskip-0.75cm m_0^2 \beta^2 \! - \! 8B \!+\! \left[ (m_0 \beta^2 \!+\! 24A)B \!+\!2 m_0^2 \beta^2 A \right] \!n^* \!+\!\left[ \frac{1}{4} B^2
\beta^2 \! - \! (m_0 \beta^2 \! + \! 48A)AB \!+\! m_0^2 \beta^2 A^2\right]\! n^{*^2} \nonumber \\
&& \hskip2.50cm + \left[ A B^2 \beta^2 + (2 m_0 \beta^2 - 16 A) A^2 B \right] n^{*^3} + A^2 B^2 \beta^2 n^{*^4} = 0.
\e
For example, for $\beta = 0.02$, one has $N_0^* = 19.70$, while for $\beta = 0.05$, the value of $N_0^*$ is $9.18$. \\
Given that to the left of $\widetilde{n}^*$ one has $L_{21}^* > 0$, the eigenvalues will have positive real parts (Figure 2b). Such critical points
are unstable and the trajectories near them are unwinding spirals (Figures 4 and 5). \\
When $N_0^* < n^* < \widetilde{n}^* = (\sqrt{3/2} - 1)/A = 22.47$ and $\beta < \sqrt{8B} / m_0 = 0.09$, the eigenvalues are both real and positive
(Figure 2b). The critical points are unstable nodes (Figures 4 and 5). When $n^* > (\sqrt{3/2} - 1)/A$, the eigenvalues are both real --- one positive
and one negative (Figure 2b) and one has saddles (Figures 4 and 5). \\
When $\beta > \sqrt{8B} / m_0 = 0.09$, the eigenvalues $\lambda_{1,2}^*$ are both real and positive for $0 < n^* < \widetilde{n}^* = (\sqrt{3/2} -
1)/A = 22.47$ (Figure 2c). These critical points are unstable nodes (Figure 6). And, finally, for $n^* > \widetilde{n}^* = (\sqrt{3/2} - 1)/A$, the
eigenvalues are both real with $\lambda_1^*$ being positive and $\lambda_2^*$ -- negative (Figure 2c). Such critical points are saddles (Figure 6). \\
The difference between the cases of $\alpha = 2$ and $\alpha > 2$ is in the 22-component ($\partial f_2 / \partial H$) of the stability matrix $L$. At
the critical point $(n^*, H^* = 0)$, it is not zero when $\alpha = 2$ and zero when $\alpha > 2$. Consider next the $\alpha = 4$ dynamical system and
denote the stability matrix by $L^{(4)}$ in this case. One has $ L^{(4)^*}_{22} = 0$ and the eigenvalues at the critical points $(n^*, H^* =0)$ are
given by
\b
\lambda_{1,2}^{(4) *} = \pm \sqrt{\frac{Bn*}{An^* + 1} \left( A^2 n^{*^2} + 2An^* - \frac{1}{2} \right)}
\e
The eigenvalues are purely imaginary, $\lambda_{1,2}^{(4) *} = \pm i \omega$, when $A^2 n^{*^2} + 2An^* - 1/2 < 0$. That is, for $n^*$ from zero to
$(\sqrt{3/2} - 1)/A$ --- exactly the point $\widetilde{n}^*$ at which $T(n^*) = T^*(n^*)$ when $\tau = \widetilde{\tau} = \sqrt{2/3}
(\sqrt{3/2}-1)^{1/3} e^{2/3-\sqrt{2/3}} A^{-1/3} B$. \\
For values of $n^*$ above $\widetilde{n}^* = (\sqrt{3/2} - 1)/A$, the eigenvalues are purely real: $\lambda_{1,2}^{(4) *} = \pm q$. \\
For $\tau > \widetilde{\tau}$, the curves $T(n^*)$ and $T^*(n^*)$ intersect only at the origin, thus critical points $(n^*, H^* = 0)$ do not exists
(see Figure 2a). \\
For $\tau < \widetilde{\tau}$, the curves $T(n^*)$ and $T^*(n^*)$ intersect, except at the origin, at points $\nu_{1,2}^*$ (see Figure 2a again) and
the intersection points $\nu_{1,2}^*$ are on either side of $\widetilde{n}^*$. Thus, at $n^* = \nu_1^*$, the eigenvalues $\lambda_{1,2}^{(4) *}$ are
purely imaginary while, at $n^* = \nu_2^*$, they are purely real (with opposite signs) and the corresponding critical points are saddles. \\
The behaviour of the trajectories near the critical points $(n^*, H^* = 0)$ for which the eigenvalues are purely imaginary, namely, for
$n^* < (\sqrt{3/2} - 1)/A$, are studied with the help of centre-manifold theory \cite{aaa} in the Appendix. One finds that all critical points with purely imaginary eigenvalues are unstable --- the trajectories near them are unwinding spirals \cite{aaa} --- see Figures 7a and 7c. \\
The origin is also a critical point. The analysis of its behaviour is done by expanding the dynamical equations near the origin and retaining only the
leading terms. For any $\alpha \ge 2$, one has:
\b
\dot{n} & = & - 3nH + 3 \beta n H^{\alpha}, \\
\dot{H} & = & -\frac{3}{2} H^2 - \frac{1}{2} \tau n^{\frac{5}{3}} + \frac{1}{2} \beta m_0 n H^{\alpha - 1} + \frac{1}{2} Bn^2 + ... \, .
\e
Consider again the separatrix $3H^2 - \rho = 0$, i.e. the second integral given by $3H^2 - n \left[ m_0 + (3/2) \tau n^{2/3} e^{2An/3} \right] + Bn^2
= 0$. Along the separatrix and near the origin, one has $3H^2 = m_0 n \,\, + $ smaller terms. Then, the equations of the dynamical system in terms of
powers of $H$ not higher than 3, reduce to $\dot{n}  =  - 3nH$ and $\dot{H} = - (3/2) H^2$. The solutions are:
\b
n(t) & = & \frac{n_0}{\bigl[ 1 + \frac{1}{2} \sigma \sqrt{3 m_0 n_0}(t-t_0)\Bigr]^2}, \\
H(t) & = & \frac{H_0}{1 + \frac{3}{2} H_0 (t - t_0)},
\e
where $\sigma = $ sgn $\!\!(H_0)$. \\
In view of the continuity, the behaviour of the trajectories near the separatrix will be the same as the behaviour along the separatrix. For the
trajectories in the upper half-plane, one will therefore have $n(t) \simeq 1/t^2$, while for those in the lower half-plane, $n(t)$ will increase with
time. Similarly, $H$ will decay to zero ($H \simeq 1/t$) for trajectories in the upper half-plane or $H$ will decrease with time for trajectories in
the lower half-plane. \\
The origin will attract trajectories from the upper half-plane and repel those from the lower half-plane. \\
There are other critical points for the $\alpha \ge 2$ dynamical system $\dot{n} = - 3 n H (1 - \beta H^{\alpha - 1}), \,\,$  $\dot{H} = - (3/2) H^2 -
(1/2) (1 - \beta H^{\alpha - 1}) p[n, T(n)]  + (1/2) \beta  H^{\alpha - 1} \rho[n, T(n)]$. \\
Clearly, if $1 - \beta H^{\alpha - 1} = 0$, then $\dot{n} = 0$ immediately and for the points $(n^{**}, H^{**})$ of the separatrix $3 H^{**^2} =
\rho(n^{**})$, for which $H^{**} = \beta^{\frac{1}{1-\alpha}}$, one will also have $\dot{H} = 0$, provided that $n^{**}$ are the solutions of $m_0
n^{**} + (3/2) n^{**} T(n^{**}) - B n^{**^2} - 3 \beta^{\frac{2}{1-\alpha}} = 0$ which can be written as:
\b
T(n^{**}) = T^{**}(n^{**})
\e
with
\b
T^{**}(n^{**}) = \frac{2}{3} (B n^{**} - m_0) + \frac{2 \beta^{\frac{2}{1-\alpha}}}{n^{**}}.
\e
Thus, such $(n^{**}, H^{**}= \beta^{\frac{1}{1-\alpha}})$ are critical points for the $\alpha \ge 2$ dynamical system, in addition to the critical
points $(n^{*}, H^{*}=0)$ and the origin. For these critical points one has:
\b
\rho^{**} \equiv \rho [ n^{**}, T^{**}(n^{**}) ] = 3 \beta^{\frac{2}{1-\alpha}}
\e
and this is greater than zero for all $n^{**}$. \\
\begin{figure}[!ht]
\centering
\subfloat[\scriptsize For $\beta < \beta_0 = (12B/m_0^2)^{\frac{\alpha-1}{2}}$, the graph of $T^{**}(n^{**})$ is entirely above the $n^{**}$-axis.
When $\beta = \beta_0$, then $T^{**}(n^{**})$ is tangent to the $n^{**}$-axis at point $n_0^{**} = \beta^{\frac{1}{1-\alpha}}\sqrt{3/B} = m_0/(2B)$.
When $\beta > \beta_0$, the function $T^{**}(n^{**})$ has zeros given by $\nu_{1,2}^{**} = \bigl(m_0/(2B)\bigr) \bigl( 1 \pm (1 - 12Bm_0^{-2}
\beta^{\frac{2}{1-\alpha}})^{\frac{1}{2}} \bigr)$ and these are equidistant from $n_0^{**}$. For any $\tau$ and $\beta$, there always exists an
intersection point $\hat{\nu}_0^{**} < \nu_{1,2}^{**}$ between the curves $T(n^{**})$ and $T^{**}(n^{**})$. Depending on $\tau$ and $\beta$, this could be the only intersection point
between the curves $T(n^{**})$ and $T^{**}(n^{**})$ or there can be one additional intersection point or two additional intersection  points between
these two curves.  Depicted here is the intersection point $\hat{\nu}_0^{**}$ between $T(n^{**})$ and $T^{**}(n^{**})$ that always exists. On Figure 3a, curve
$T(n^{**})$ with fixed $\tau$ is chosen and it intersects curves $T^{**}(n^{**})$ with varying $\beta$. See Figure 3b for the remaining intersection
points --- when they exist, they are at higher  $n^{**}$.]
{\label{F3a}\includegraphics[height=4.1cm, width=0.33\textwidth]{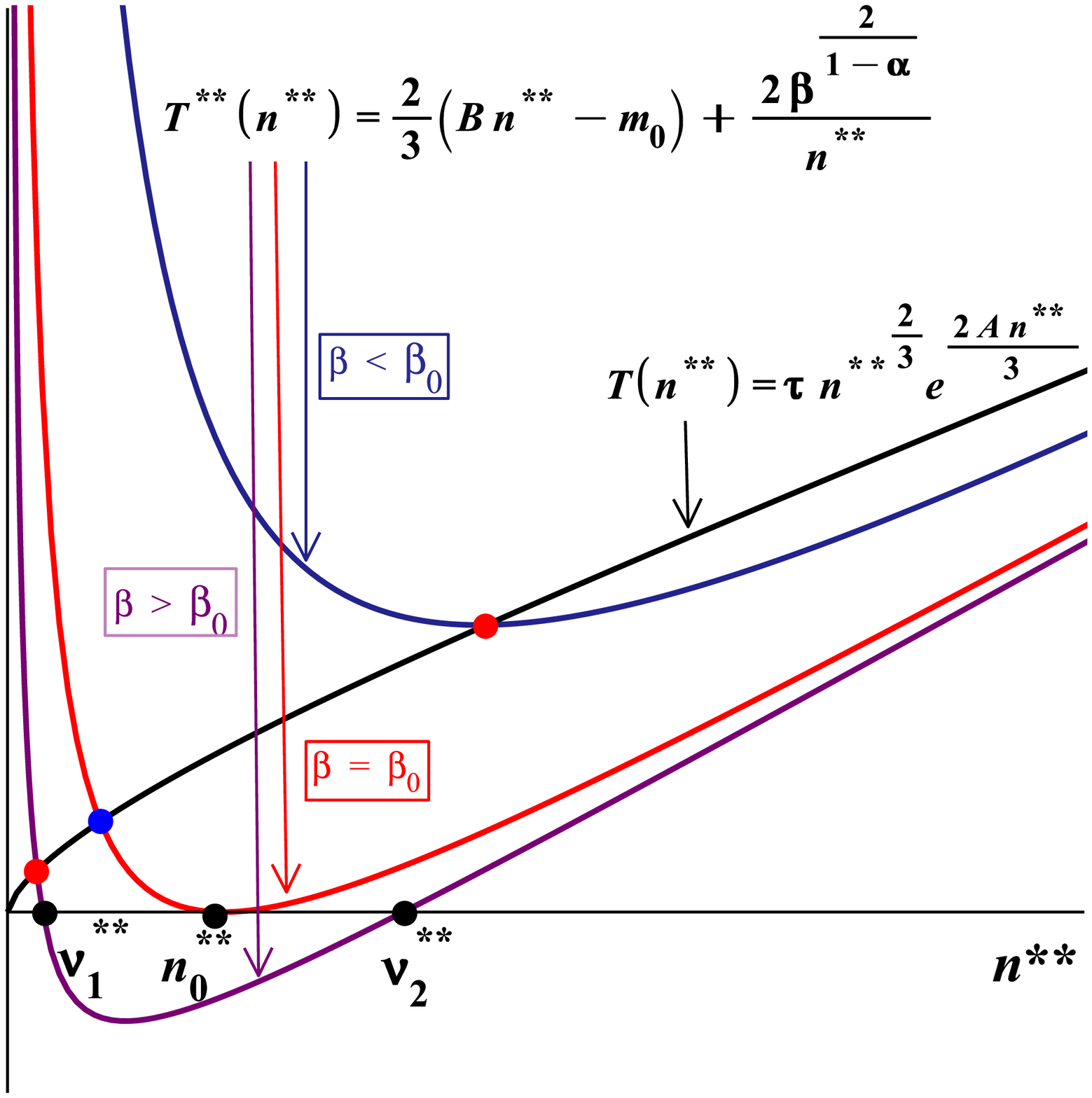}}
\,
\subfloat[\scriptsize The number of intersection points of $T(n^{**})$ with $T^{**}(n^{**})$ and their loci depend on $\tau$ and $\beta$. Taken here
is curve $T^{**}(n^{**})$ with $\beta > \beta_0$, any other choices of $\beta$ are treated in an entirely analogical manner. The curves $T(n^{**})$
are taken with varying $\tau$. For $\tau = \hat{\tau}$, the curves $T(n^{**})$ and $T^{**}(n^{**})$ are tangent to each other at point $\hat{n}^{**}$,
where $\hat{n}^{**}$ and $\hat{\tau}$ are solutions to (\ref{nhat}) and (\ref{tauhat}) respectively. When $\tau > \hat{\tau}$, the curves $T(n^{**})$
and $T^{**}(n^{**})$ do not intersect elsewhere, except at the point shown on Figure 3a. When $\tau < \hat{\tau}$, then $T(n^{**})$ and
$T^{**}(n^{**})$ intersect at points $\hat{\nu}_{1,2}^{**}$ (which are greater than $\nu_{1,2}^{**}$ when $\nu_{1,2}^{**}$ exist, that is, when  $\beta > \beta_0$) --- additional to the intersection point $\hat{\nu}_0^{**}$ shown on Figure 3a. Point $\hat{\nu}_{1}^{**}$ is to the left of $\hat{n}^{**}$, while point $\hat{\nu}_{2}^{**}$ is to the right of $\hat{n}^{**}$. For the numerical example considered, one has $\hat{n}^{**} = 134.33$ and $\hat{\tau} = 12.93$ when $\alpha = 2$ and $\hat{n}^{**} = 134.81$ and $\hat{\tau} = 12.88$ when $\alpha = 4$.]
{\label{F3b}\includegraphics[height=4.1cm,width=0.32\textwidth]{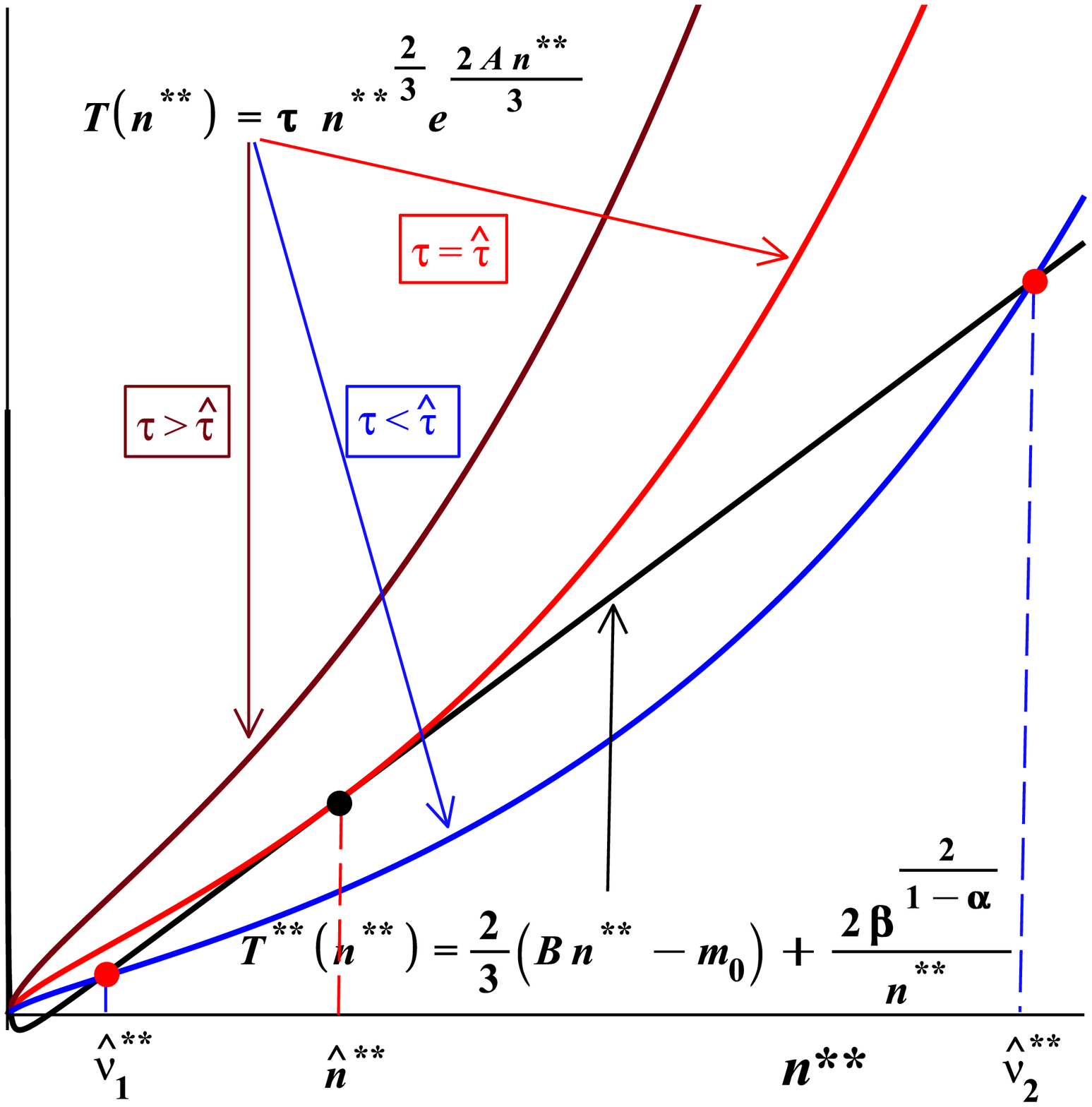}}
\,
\subfloat[\scriptsize For $\alpha \ge 2$, critical points $(n^{**}, H^{**} = \beta^{\frac{1}{1-\alpha}})$ are stable if $L_{21}^{**}$ is negative,
that is, when $T^{**}(n^{**}) < Q(n^{**}) = (2 B n^{**} - m_0)/(A n^{**} + 5/2)$. When $\beta > \beta_0$, curve {\bf (i)}, $T^{**}(n^{**})$,
intersects the $n^{**}$-axis at points $\nu_{1,2}^{**}$ and it also intersects the curve $Q(n^{**})$ at points $\sigma_{1,2}^{**}$. Critical points
with $\nu_{2}^{**} < n^{**} < \sigma_2^{**}$ are stable (note that there can be no critical points of this type where $T^{**}(n^{**})$ is negative).
When $\beta = \beta_0$, curve {\bf (ii)} is tangent to the $n^{**}$-axis at point $\chi_{1}^{**} = n_0^{**}$ --- the point at which $Q(n^{**})$
crosses the abscissa. Further, {\bf (ii)} intersects the curve $Q(n^{**})$ at point $\chi_{2}^{**}$ and critical points with $\chi_{1}^{**} < n^{**} <
\chi_2^{**}$ are stable. Curve {\bf (iii)} is characterised by $\beta_Q < \beta < \beta_0$. This curve never intersects the $n^{**}$-axis and it
intersects the curve $Q(n^{**})$ at $\xi_{1,2}^{**}$. Critical points with $\xi_{1}^{**} < n^{**} < \xi_2^{**}$ are stable. Finally, curve {\bf (iv)}
is characterised by $\beta < \beta_Q$. This curve never intersects the $n^{**}$-axis or the curve $Q(n^{**})$. There are no stable critical points in
this case.]
{\label{F3c}\includegraphics[height=4.1cm,width=0.32\textwidth]{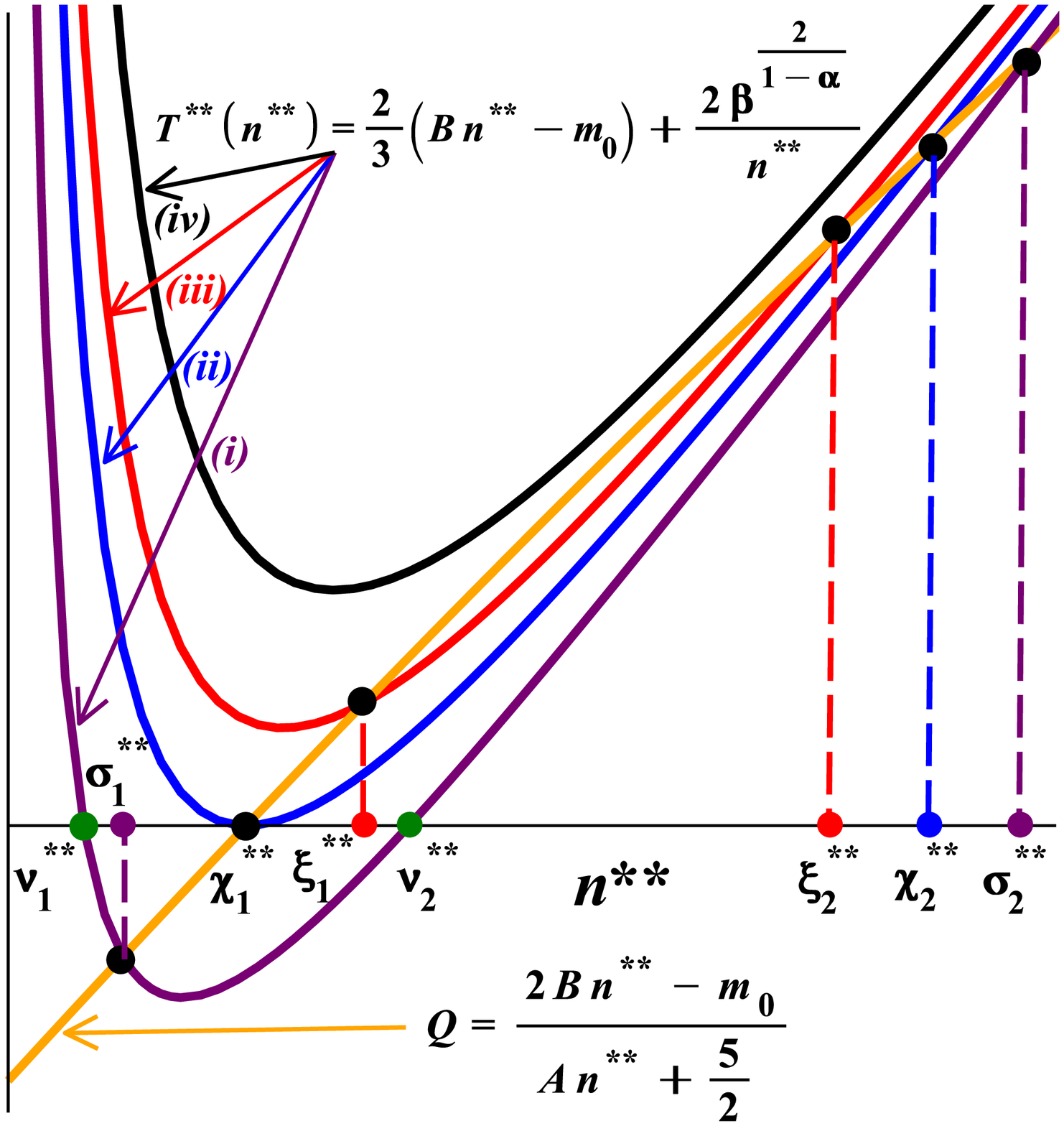}}
\caption{\footnotesize{Determination of the critical points of the type $(n^{**}, H^{**} = \beta^{\frac{1}{1-\alpha}})$ for the $\alpha \ge 2$
dynamical system. The loci $n^{**}$ of the critical points are the solutions of $T^{**}(n^{**}) = T(n^{**})$ --- Figures 3a and 3b. Figure 3c shows
where stable critical points of the type $(n^{**}, H^{**} = \beta^{\frac{1}{1-\alpha}})$ can be found. }}
\label{Figure3}
\end{figure}
\n
$\!\!\!$Since the critical points $(n^{**}, H^{**})$ are on the separatrix, one should solve the equation for the trajectory reaching or moving away from such
critical point firstly while on the separatrix itself. Substituting $\rho = 3H^2$ into the dynamical equation for $H$ yields:
\b
\dot{H} = -\frac{1}{2} (1 - \beta H^{\alpha - 1}) \left[ 3H^2 + p[n,T(n)] \right]
\e
and then, expanding about $H^{**}$, gives:
\b
\dot{H}= \frac{1}{2}\beta (\alpha-1)H^{**^{\alpha - 2}}[3H^{**^2}+ p^{**}](H - H^{**}) = \kappa (H - H^{**}),
\e
where $\kappa = (1/2)\beta (\alpha-1)H^{**^{\alpha - 2}}[3H^{**^2}+ p^{**}] \!=\! (1/2)\beta (\alpha-1)H^{**^{\alpha - 2}}
[ (5 + 2 A n^{**})\beta^{\frac{2}{1-\alpha}}$ \linebreak $ + (2 n^{**} / 3)(1 + A n^{**})(B n^{**}-m_0) - Bn^{**^2}] = $ const. \\
The solution along the separatrix near the critical point $(n^{**}, H^{**})$ is therefore:
\b
\ln \left| \frac{H-H^{**}}{H_0-H^{**}} \right| = \kappa (t-t_0).
\e
The sign of $\kappa$ is important. When $\kappa > 0$, in order to get $H \to H^{**}$, it is necessary to have $t \to -\infty,$ i.e. the separatrix in
this case is an unstable curve of a saddle or the critical point is an unstable node. When $\kappa < 0$, one has $H \to H^{**}$ as $t \to \infty,$
i.e. the separatrix in this case is a stable curve of a saddle or the critical point is a stable node. In view of the continuity, trajectories close
to the separatrix will exhibit similar behaviour. \\
The function $T^{**}(n^{**})$  has a minimum at $\beta^{\frac{1}{1-\alpha}}\sqrt{3/B}$. When $\beta$ equals  $\beta_0 \! = \!
(12B/m_0^2)^{\frac{\alpha-1}{2}}\!$, this minimum will occur at $n_0^{**}$ from the $n^{**}$-axis: $n_0^{**} = \beta^{\frac{1}{1-\alpha}}\sqrt{3/B} =
m_0/(2B)$. For values of $\beta < \beta_0$, the graph of $T^{**}(n^{**})$ is entirely above the $n^{**}$-axis, while for $\beta > \beta_0$, the
function $T^{**}(n^{**})$ has zeros given by $\nu_{1,2}^{**} = [m_0/(2B)] \, [ 1 \pm (1 - 12Bm_0^{-2} \beta^{\frac{2}{1-\alpha}})^{\frac{1}{2}} ]$ ---
see Figure 3a. When $\alpha = 2$, for the numerical example considered one has $\beta_0 = 0.1095$, while for $\alpha = 4,$ the corresponding value is
$\beta_0 = 0.0013$.  \\
Depending on the parameters $\beta$ and $\tau$, the number of intersection points of the curves $T(n^{**})$ and $T^{**}(n^{**})$ is one, two, or three
--- see Figures 3a and 3b. At some value $\hat{\tau}$ of $\tau$, for any given $\beta$, the curves $T(n^{**})$ and $T^{**}(n^{**})$ are tangent to
each other at point, say $\hat{n}^{**}$. At this point, the tangents to the two curves coincide, thus one has the following two simultaneous
equations: $T(\hat{n}^{**}) = T^{**}(\hat{n}^{**})$ and $(d/dn^{**}) [T(n^{**})]_{(n^{**} = \hat{n}^{**}, \tau = \hat{\tau})} = (d/dn^{**})
[T^{**}(n^{**})]_{(n^{**} = \hat{n}^{**}, \tau = \hat{\tau})}.$
The solution of this system is $\hat{n}^{**}$, which satisfies
\b
\label{nhat}
2 A B \hat{n}^{**^3} - (2 m_0 A + B)\hat{n}^{**^2} + (6 A \beta^{\frac{2}{1-\alpha}} - 2m_0) \hat{n}^{**} + 15 \beta^{\frac{2}{1-\alpha}} = 0,
\e
and $\hat{\tau}$ given by
\b
\label{tauhat}
\hat{\tau} = \left[ \frac{2}{3} (B \hat{n}^{**} - m_0) + \frac{2 \beta^{\frac{2}{1-\alpha}}}{\hat{n}^{**}} \right] \hat{n}^{**^{-\frac{2}{3}}}
e^{-\frac{2A\hat{n}^{**}}{3}}.
\e
When $\tau < \hat{\tau}$, that is, when points $\hat{\nu}_{1,2}^{**}$  exist, one has $\hat{\nu}_{1}^{**}$ to the left of $\hat{n}^{**}$ and
$\hat{\nu}_{2}^{**}$ to the right of $\hat{n}^{**}$. \\
The components of the stability matrix $L$ at the critical points $(n^{**}, H^{**}= \beta^{\frac{1}{1-\alpha}})$ are: $L^{**}_{11} = 0, \,\,
L^{**}_{12} = 3 (\alpha - 1) n^{**}$,
\b
L^{**}_{21} & = & \frac{1}{2} T^{**}(n^{**}) (A n^{**} + \frac{5}{2}) - B n^{**} + \frac{m_0}{2} \nonumber \\
&& = \frac{1}{3} [A B n^{** ^2} - (m_0 A + \frac{B}{2}) n^{**} - m_0 ] + \frac{A}{\beta^{\frac{2}{\alpha - 1}}}
+ \frac{5}{2 \beta^{\frac{2}{\alpha - 1}} n^{**}}, \\
L^{**}_{22} & = & - 3\beta^{\frac{1}{1 - \alpha}} \, \, + \,\, \frac{1}{2} \, (\alpha - 1) \, \beta^{\frac{1}{\alpha - 1}} \,
n^{**} \, [ T^{**}(n^{**}) (A n^{**} + \frac{5}{2}) - 2 B n^{**} + m_0] \nonumber \\
&& = - 3\beta^{\frac{1}{1 - \alpha}} \,\, + \,\, (\alpha - 1) \, \beta^{\frac{1}{\alpha - 1}} \, n^{**} \, L^{**}_{21}.
\e
The eigenvalues are always real:
\b
\lambda_1^{**} & \! = \! & - 3\beta^{\frac{1}{1 - \alpha}} < 0, \\
\label{la2}
\lambda_2^{**} & \! = \! & (\alpha - 1) \, \beta^{\frac{1}{\alpha - 1}} \, n^{**} \, L^{**}_{21} = \frac{1}{2} \, (\alpha - 1) \,
\beta^{\frac{1}{\alpha - 1}} \,
n^{**} \, [ T^{**}(n^{**}) (A n^{**} + \frac{5}{2}) - 2 B n^{**} + m_0]. \nonumber \\
\e
Given that $\lambda_1^{**} < 0$,  the critical points $(n^{**}, H^{**} = \beta^{\frac{1}{1-\alpha}})$ will be stable if $\lambda_2^{**}$ is negative,
that is, if $L^{**}_{21} < 0$ or if
\b
T^{**}(n^{**}) < Q(n^{**}) \equiv \frac{2 B n^{**} - m_0}{A n^{**} + \frac{5}{2}}.
\e
Otherwise, the critical points $(n^{**}, H^{**} = \beta^{\frac{1}{1-\alpha}})$ will be saddles. \\
Four curves $T^{**}(n^{**})$ with different $\beta$ are shown on Figure 3c, together with the curve $Q(n^{**})$ which starts at point $(0, -2m_0/5)$,
crosses the $n^{**}$-axis at $n_0^{**} = m_0/(2B)$ and has a horizontal asymptote at $2B/A$. When $\beta > \beta_0= (12B/m_0^2)^{(\alpha - 1)/2}$, the
curve $T^{**}(n^{**})$, marked with {\bf (i)} on Figure 3c, intersects the $n^{**}$-axis at points $\nu_{1,2}^{**}$. The $n^{**}$-coordinates of the
intersection point of $T^{**}(n^{**})$ with the curve $Q(n^{**})$ are $\sigma_{1,2}^{**}$. As, while negative, $T^{**}(n^{**})$ cannot intersect the
strictly positive $T(n^{**})$, no critical points $(n^{**}, H^{**} = \beta^{1/(1-\alpha)})$ can exist for $T^{**}(n^{**}) < 0$. Thus, stable critical
points for $\beta > \beta_0$ exist in the interval $\nu_{2}^{**} < n^{**} < \sigma_2^{**}$ ---  where the non-negative $T^{**}(n^{**})$  is smaller
than $Q(n^{**})$. When $\beta = \beta_0$, the curve $T^{**}(n^{**})$, marked with {\bf (ii)} on Figure 3c, is tangent to the $n^{**}$-axis at point
$\chi_{1}^{**} = n_0^{**}$ --- the point at which $Q(n^{**})$ crosses the abscissa. This curve intersects the curve $Q(n^{**})$ further --- at point
$\chi_{2}^{**}$. Critical points for which $\chi_{1}^{**} < n^{**} < \chi_2^{**}$ are stable. \\
There is a value of $\beta$, say $\beta_Q$, for which, at certain $n_Q^{**}$ from the $n^{**}$-axis, the curve $T^{**}(n^{**})$ is tangent to the
$\beta$-independent curve $Q(n^{**})$. That is, at $n_Q^{**}$, the two functions are equal, $T^{**}(n_Q^{**}) = Q(n_Q^{**})$, and their first
derivatives are also equal, $(d/dn^{**}) [T(n^{**})]_{(n^{**} = n_Q^{**}, \, \beta = \beta_Q)} = (d/dn^{**}) [Q(n^{**})]_{(n^{**} = n_Q^{**}, \, \beta
= \beta_0)}.$ Thus, $n_Q^{**}$ is found, for any $\alpha \ge 2$,  as the only positive root of the cubic equation
\b
4 A^2 B n_Q^{**^3} - 2 A (m_0 A - 7 B) n_Q^{** ^2} - 5 (2 m_0 A + B) n_Q^{**} - 5 m_0 = 0.
\e
For the numerical example considered, one gets $n_Q^{**} = 45.4587$ and, hence, $\beta_Q = 0.03426$ for $\alpha = 2$ and $\beta_Q = 0.00004$ for
$\alpha = 4$. \\
When $\beta_Q < \beta < \beta_0$, curve $T^{**}(n^{**})$,  marked with {\bf (iii)} on Figure 3c, never intersects the $n^{**}$-axis. It intersects the
curve $Q(n^{**})$ at points with $n^{**}$ coordinates given by $\xi_{1,2}^{**}$. Critical points with $\xi_{1}^{**} < n^{**} < \xi_2^{**}$ are stable.
Finally, when $\beta < \beta_Q$, curve $T^{**}(n^{**})$,  marked with {\bf (iv)} on Figure 3c, never intersects the $n^{**}$-axis or the curve
$Q(n^{**})$. There are no stable critical points in this case. \\
For the dynamical system in the case of $\alpha = 2$, three sub-cases are considered: $\beta = 0.02$ (Figure 4), $\beta = 0.05$ (Figure 5), and $\beta = 0.1$ (Figure 6). With these, all qualitatively different possibilities are analyzed. The case of $\alpha = 4$ is similar --- see Figure 7 where some representative cases are shown. The two Tables at the end should also be considered as all possibilities for the model parameters are summarized there and references are given to the corresponding Figures. \\
Many of the trajectories exhibit inflationary regime (Figures 4 -- 7). This happens in the upper half-plane ($H > 0$) and while $H$ is increasing ($\dot{H} > 0$), thus $\ddot{a} > 0$. The un-physical trajectories that diverge to $(n \to \infty, H \to \infty)$ have eternal inflation, while all other trajectories with inflation, after exiting their inflationary regimes, either extinguish at the origin $(n \to 0, H \to 0)$ in infinite time (Big Freeze); or at a stable critical point in infinite time; or diverge to a Big Crunch: $(n \to \infty, H \to -\infty)$.
\begin{figure}[!ht]
\centering
\subfloat[\scriptsize There are four critical points when $\alpha = 2$, $\beta = 0.02$ and $\tau = 14$: the origin, the other two intersections of
$T^*(n^*)$ with $T(n^*)$, namely the unstable critical point $(n^* = 3.22, H^* = 0)$ --- non-discernible due to the scale of this diagram --- around which trajectories spiral out and the saddle at $(n^* =
80.36, H^* = 0)$, and also the single intersection point of $T^{**}(n^{**})$ with $T(n^*)$, namely the saddle at $(n^{**}=125.40,H^{**}= 50)$.The
saddle at $(n^* = 80.36, H^* = 0)$ is with $\rho^* < 0$.]
{\label{F4a}\includegraphics[height=4.3cm, width=0.32\textwidth]{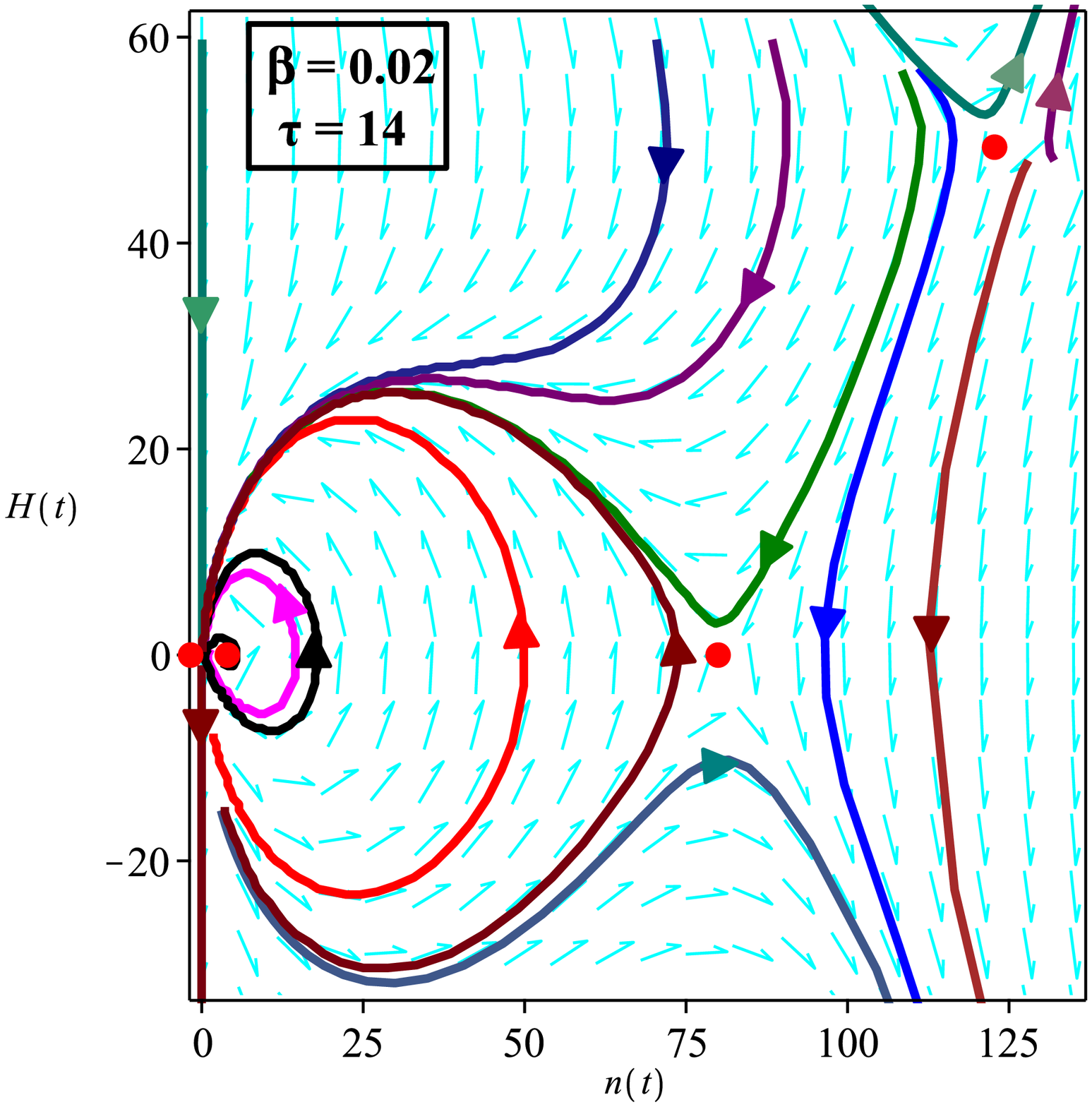}}
\,
\subfloat[\scriptsize Again, there are four critical points when $\alpha = 2$, $\beta = 0.02$ and $\tau = 18$ (as in the case on Figure 4a): the
origin, the other two intersections of $T^*(n^*)$ with $T(n^*)$, namely the unstable critical point $(n^* = 9.07, H^* = 0)$, around which trajectories
spiral out, and a saddle at $(n^* = 46.53, H^* = 0)$, and also the single intersection point of $T^{**}(n^{**})$ with $T(n^{**})$, namely the saddle
at $(n^{**}=62.50, H^{**}=50)$. The situation is similar to the one on Figure 7a, but this time the saddle at $(n^* = 46.53, H^* = 0)$ has $\rho^* >
0$.]
{\label{F4b}\includegraphics[height=4.3cm, width=0.32\textwidth]{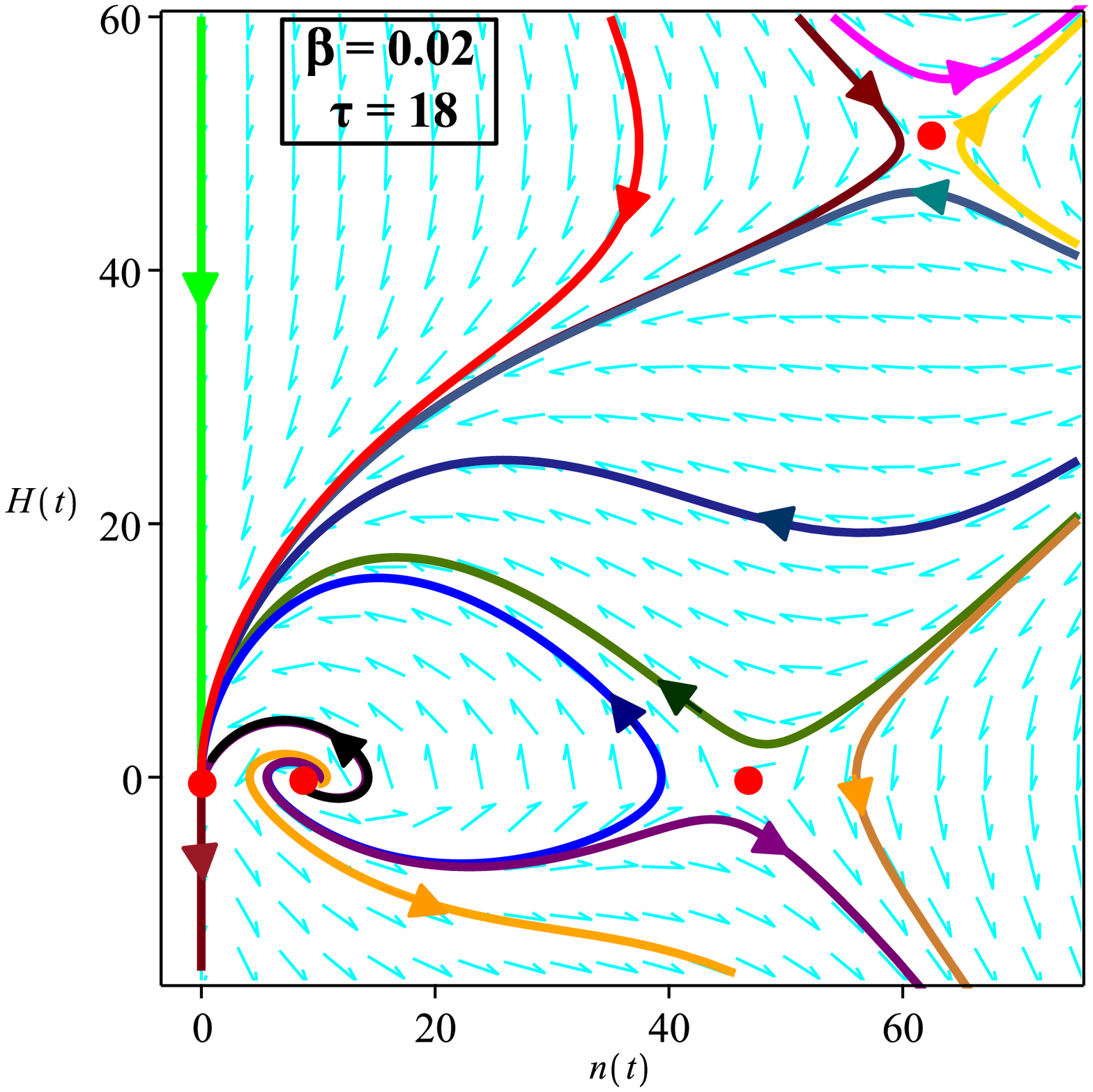}}
\,
\subfloat[\scriptsize There are two critical points when $\alpha = 2$, $\beta = 0.02$ and $\tau = 24$: the origin, which is the only intersection of
$T^*(n^*)$ with $T(n^*)$, and the single intersection point of $T^{**}(n^{**})$ with $T(n^{**})$ --- the saddle at $(n^{**} = 32.55, H^{**} = 50)$.]
{\label{F4c}\includegraphics[height=4.3cm,width=0.32\textwidth]{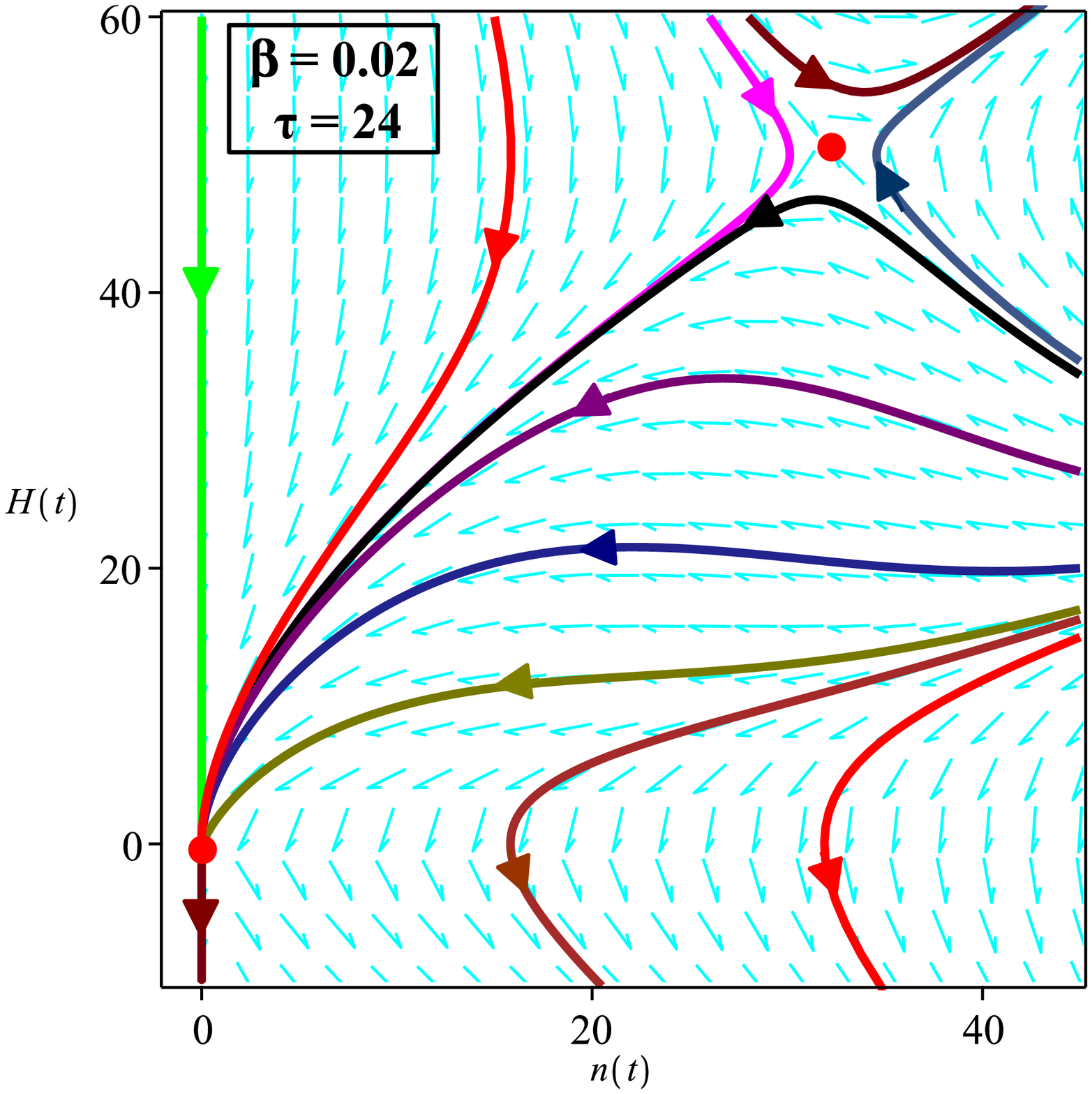}}
\caption{\footnotesize{The case of $\alpha = 2$ with $\beta = 0.02$. As $\beta < \sqrt{8B}/m_0 = 0.0894$, the eigenvalues $\lambda_{1,2}^*$ are both
complex with positive real parts for $0 < n^* < N_0^* = 19.70$. The trajectories near them are unwinding spirals (see Figures 4a and 4b). For values
of $n^*$ between $N_0^* = 19.70$ and $(\sqrt{3/2}-1)/A = 22.47$, the eigenvalues are both real and positive. The critical points are unstable nodes.
Finally, when $n^* > (\sqrt{3/2} - 1)/A = 22.47$, the eigenvalues are both real --- one positive and one negative and one has saddles. As $\beta =
0.02 < \beta_Q = 0.0343$, both eigenvalues $\lambda_{1,2}^{**}$ are real and with opposite signs for all $n^{**}$, thus the corresponding critical points are
always saddles. }}
\label{Figure4}
\end{figure}

\begin{figure}[!ht]
\centering
\subfloat[\scriptsize There are six critical points when $\alpha = 2$, $\beta = 0.05$ and $\tau = 14$: the origin, the other two intersections of
$T^*(n^*)$ with $T(n^*)$, namely the unstable critical point $(n^* = 3.22, H^* = 0)$ (shown here), around which trajectories spiral out, and a saddle
at $(n^* = 80.38, H^* = 0)$, shown on Figure 5c, and also the three intersections of $T^{**}(n^{**})$ with $T(n^{**})$, namely the two saddles
$(n^{**} = 12.13, H^{**} = 20)$ and $(n^{**} = 134.69, H^{**} = 20)$, and a stable node $(n^{**} = 34.69, H^{**} = 20)$ --- see Figure 5b and 5c for
these.  ]
{\label{F5a1}\includegraphics[height=4.3cm, width=0.32\textwidth]{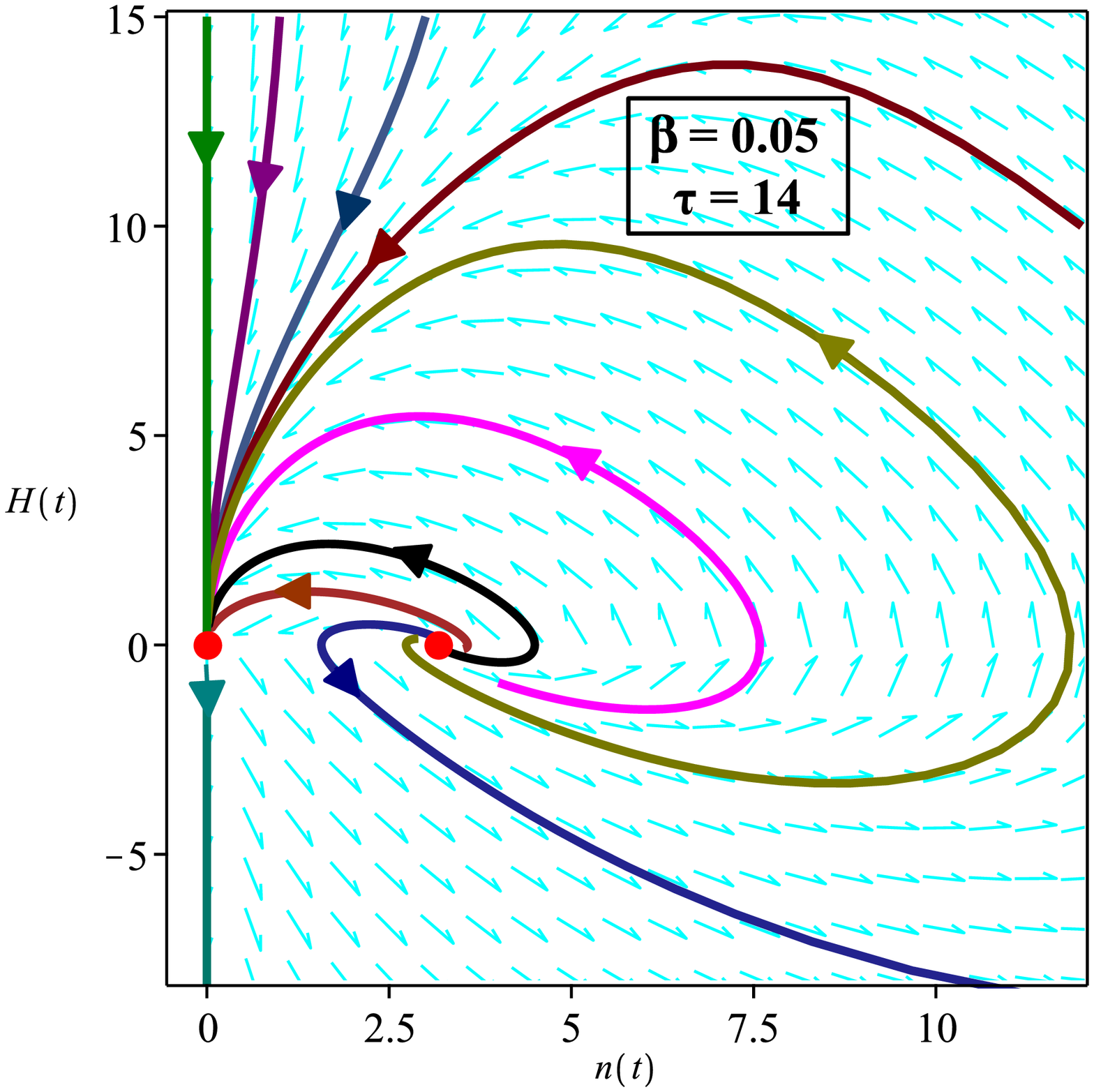}}
\,
\subfloat[\scriptsize Continuation of Figure 5a: the critical point at $(n^{**} = 12.13, H^{**} = 20)$ is a saddle, while the one at $(n^{**} = 34.69,
H^{**} = 20)$ is a stable node. ]
{\label{F5a2}\includegraphics[height=4.3cm,width=0.32\textwidth]{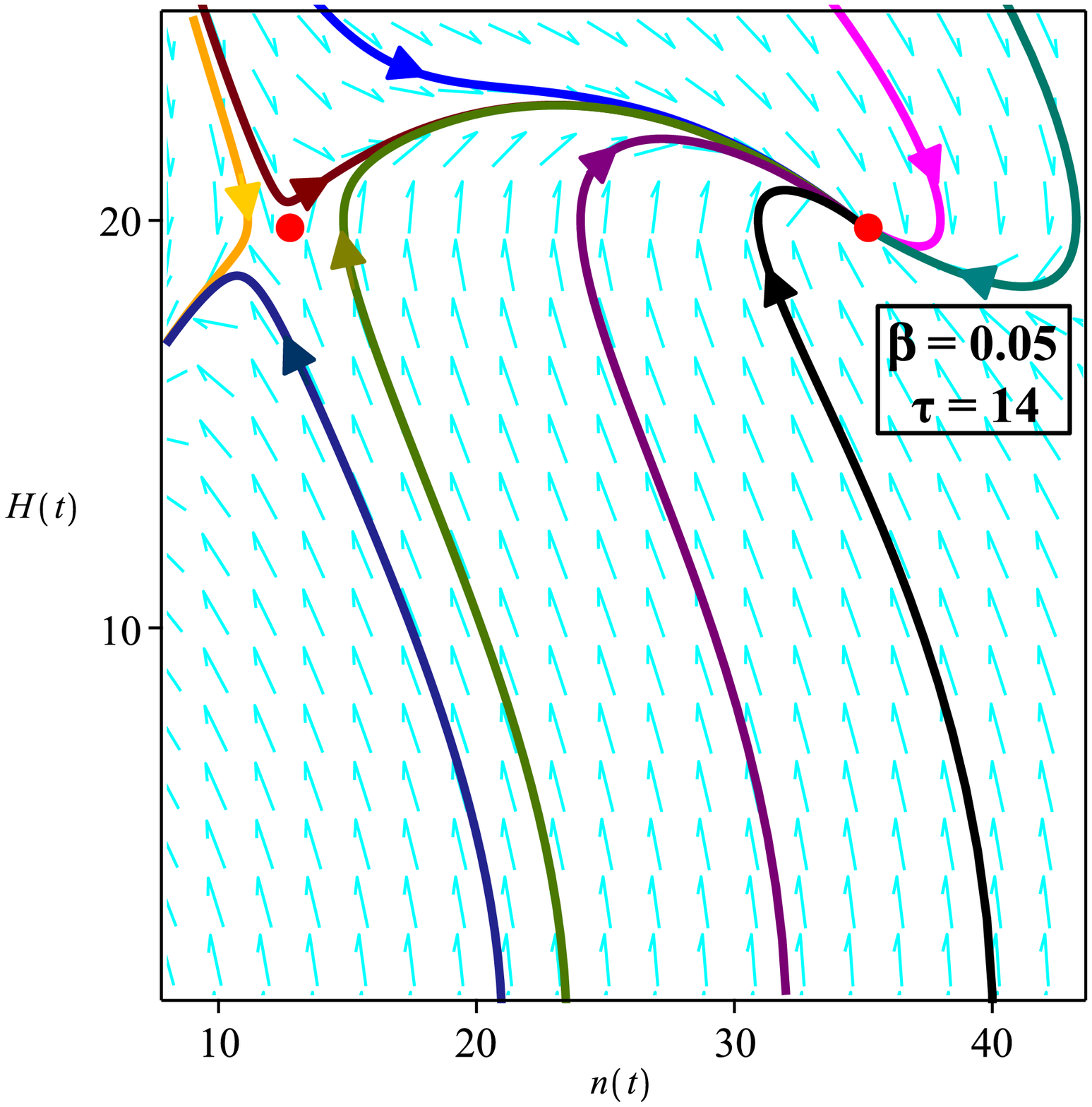}}
\,
\subfloat[\scriptsize  Continuation of Figures 5a and 5b: the critical point at $(n^{**} = 112.38, H^{**} = 20)$ is a saddle. The critical point at
$(n^* = 80.38, H^*=0)$ is also a saddle. At the latter, $\rho^* < 0$.]
{\label{F5a3}\includegraphics[height=4.3cm,width=0.32\textwidth]{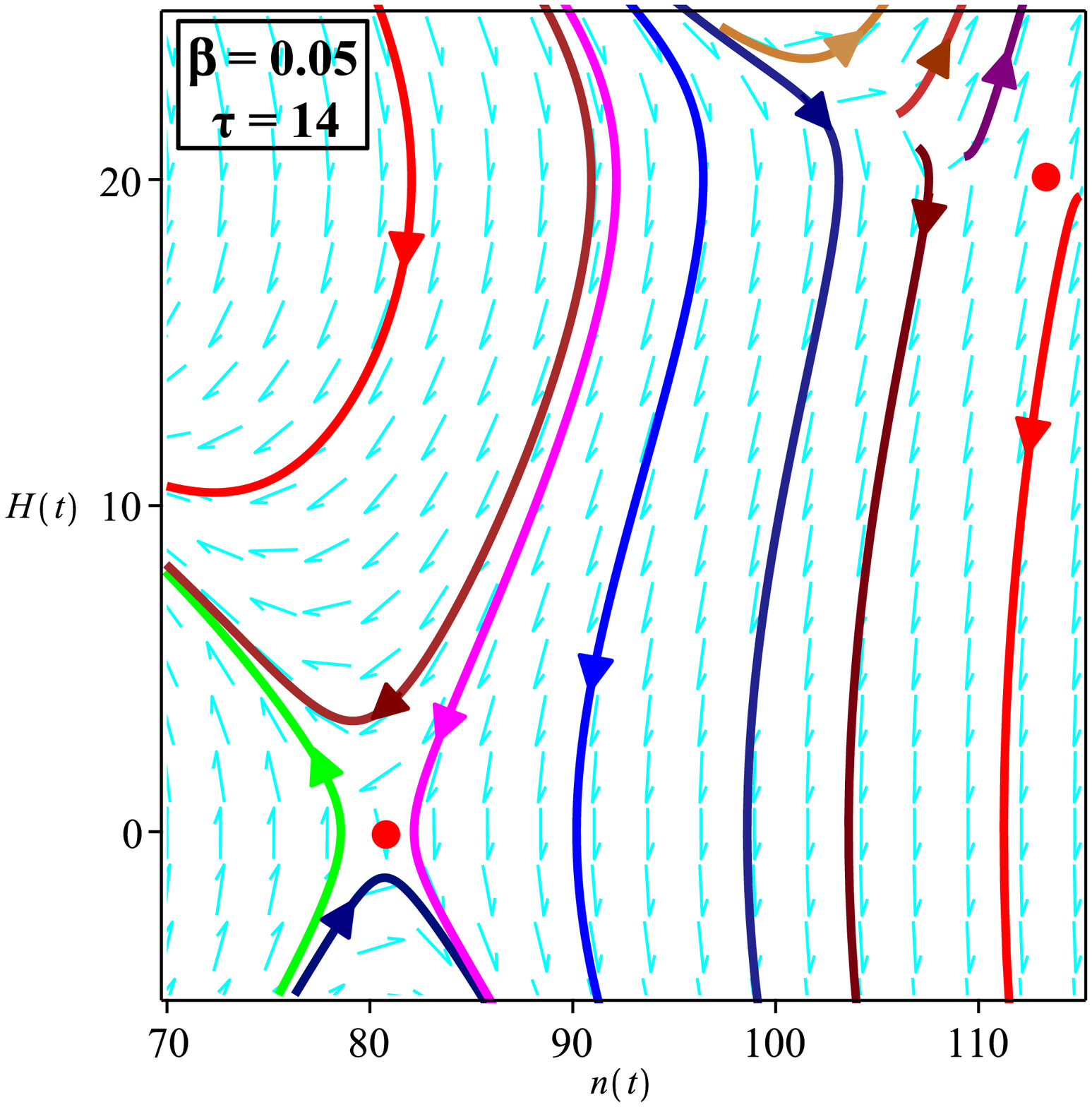}}
\caption{\footnotesize{Parts (a), (b), and (c) --- the case of $\alpha = 2$ with $\beta = 0.05$. As $\beta < \sqrt{8B}/m_0 = 0.0894$, the eigenvalues
$\lambda_{1,2}^*$ are both complex with positive real parts for $0 < n^* < N_0^* = 9.19$. The trajectories near them are unwinding spirals (see
Figures 5a and 5d). For values of $n^*$ between $N_0^* = 9.19$ and $(\sqrt{3/2}-1)/A = 22.47$, the eigenvalues are both real and positive. The
critical points are unstable nodes. Finally, when $n^* > (\sqrt{3/2} - 1)/A = 22.47$, the eigenvalues are both real --- one positive and one negative
and one has saddles (see Figure 5c). In relation to the eigenvalues $\lambda_{1,2}^{**}$, one has $n_1^{**} = 18.27$ and $n_2^{**} = 66.45$. Critical
points with $0 < n^{**} < n_1^{**} = 18.27$ are with real eigenvalues with opposite signs (saddles, see Figures 5b, 5d, and 5f), those with $n^*$
between $n_1^{**} = 18.27$ and $n_2^{**} = 66.45$ are with real and negative eigenvalues (stable nodes, see Figure 5b), and critical points with $n^*$
above $n_2^{**} = 66.45$ are with real eigenvalues with opposite signs (saddles, see Figure 5c).}}
\label{Figure5_1}
\end{figure}

\addtocounter{figure}{-1}
\addtocounter{subfigure}{+3}

\begin{figure}[!ht]
\centering
\subfloat[\scriptsize There are four critical points when $\alpha = 2$, $\beta = 0.05$ and $\tau = 18$: the origin, the other two intersections of
$T^*(n^*)$ with $T(n^*)$, namely the unstable critical point $(n^* = 9.07, H^* = 0)$ (shown here), around which trajectories spiral out, and a saddle
at $(n^* = 46.53, H^* = 0)$, shown on Figure 5e, and also the single intersection point of $T^{**}(n^{**})$ with $T(n^{**})$, namely the saddle at
$(n^{**}=8.95, H^{**}=20)$.]
{\label{F5b1}\includegraphics[height=4.3cm,width=0.32\textwidth]{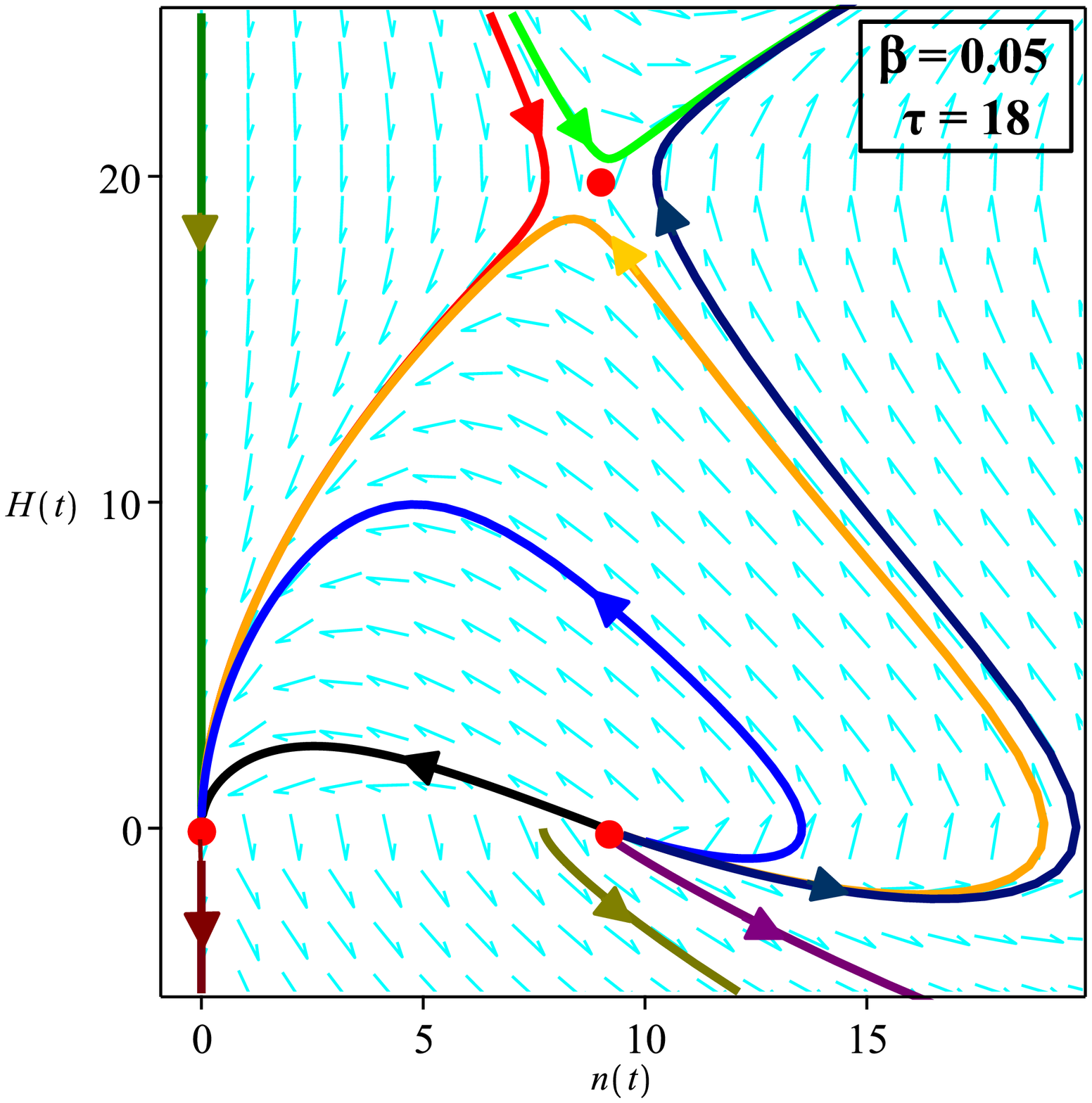}}
\,
\subfloat[\scriptsize Continuation of Figure 5d: the critical point at $(n^* = 46.53, H^* = 0)$ is a saddle.]
{\label{F5b2}\includegraphics[height=4.3cm,width=0.32\textwidth]{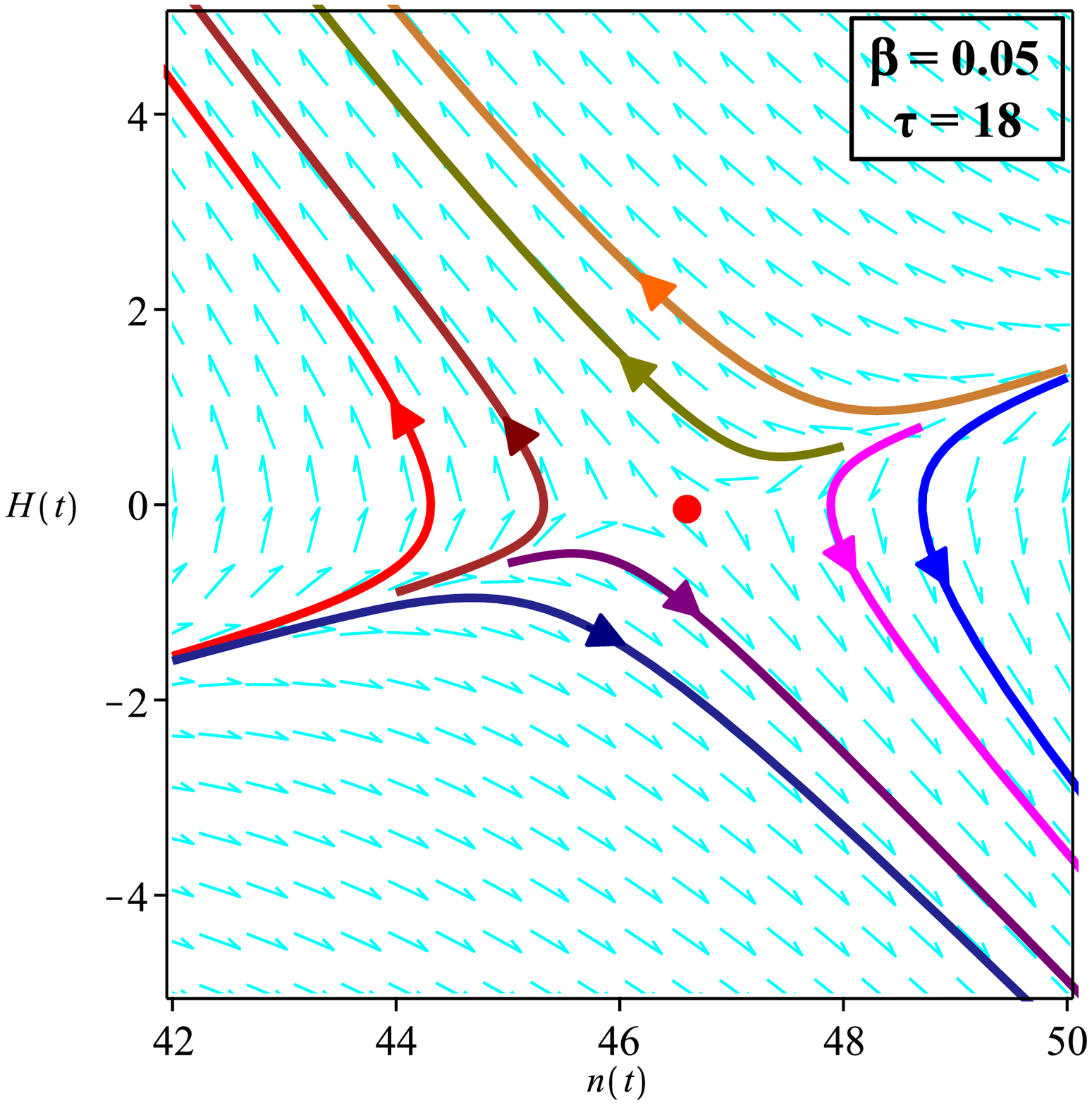}}
\,
\subfloat[\scriptsize There are two critical points when $\alpha = 2$, $\beta = 0.05$ and $\tau = 22$: the origin, which is the only intersection of
$T^*(n^*)$ with $T(n^*)$, and the single intersection point of $T^{**}(n^{**})$ with $T(n^{**})$ --- the saddle at $(n^{**} = 7.59, H^{**} = 20)$. ]
{\label{F5c}\includegraphics[height=4.3cm,width=0.32\textwidth]{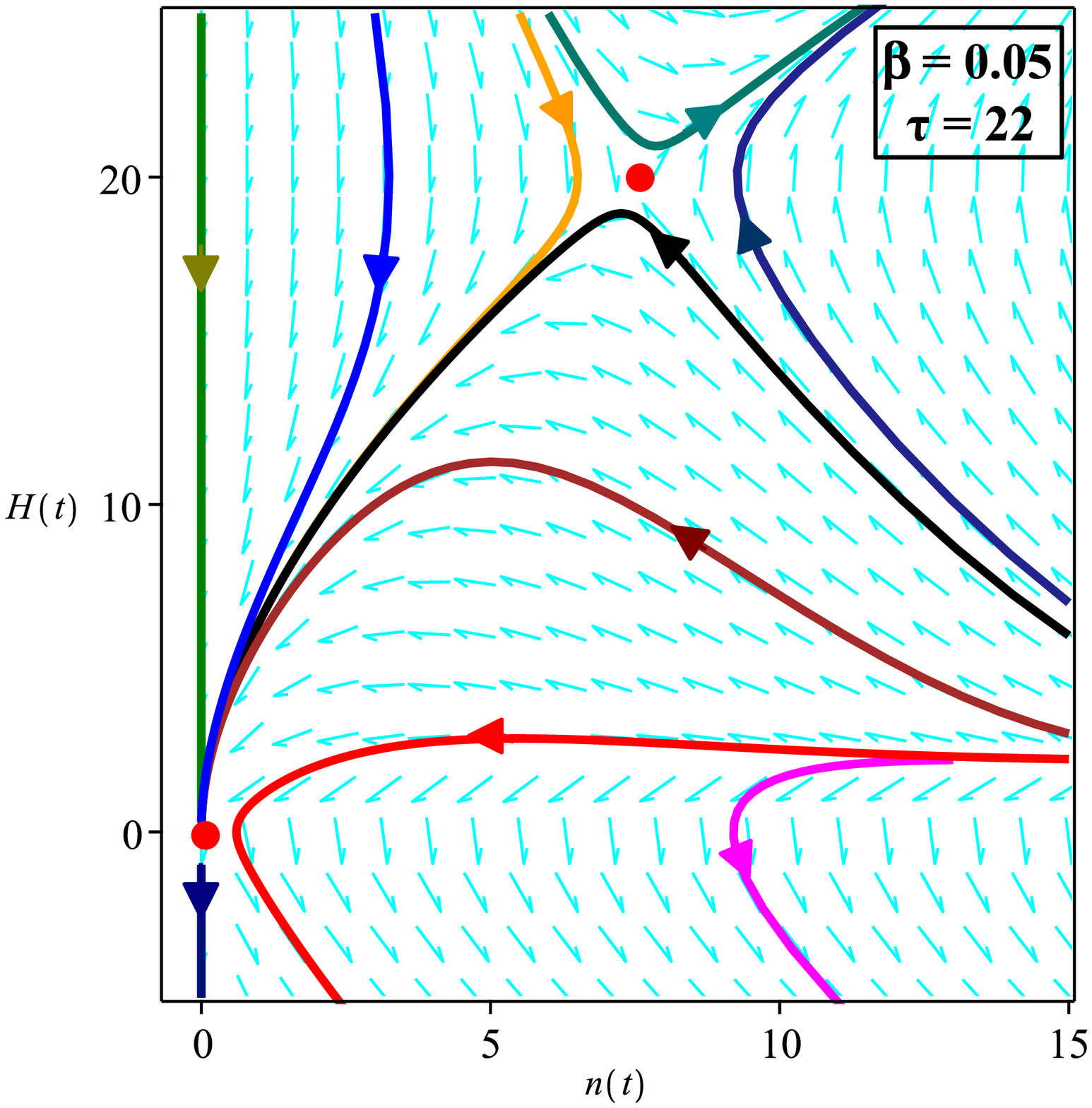}}
\caption{\footnotesize{Parts (d), (e), and (f) --- the case of $\alpha = 2$ with $\beta = 0.05$.}}
\label{Figure5_2}
\end{figure}

\begin{figure}[!ht]
\centering
\subfloat[\scriptsize There are six critical points when $\alpha = 2$, $\beta = 0.1$ and $\tau = 14$: the origin, the other two intersections of
$T^*(n^*)$ with $T(n^*)$, namely the unstable critical point $(n^* = 3.2, H^* = 0)$ (shown here) and a saddle at $(n^* = 80.4, H^* = 0)$, shown on
Figure 6b, and also the three intersections of $T^{**}(n^{**})$ with $T(n^{**})$, namely the saddle $(n^{**} =2.6, H^{**} = 10)$ (shown here), the
stable node $(n^{**} = 45.9, H^{**} = 10)$, shown on Figure 6b, and the saddle  $(n^{**} = 109.6, H^{**} = 10)$, shown on Figure 6c.]
{\label{F6a1}\includegraphics[height=4.3cm, width=0.32\textwidth]{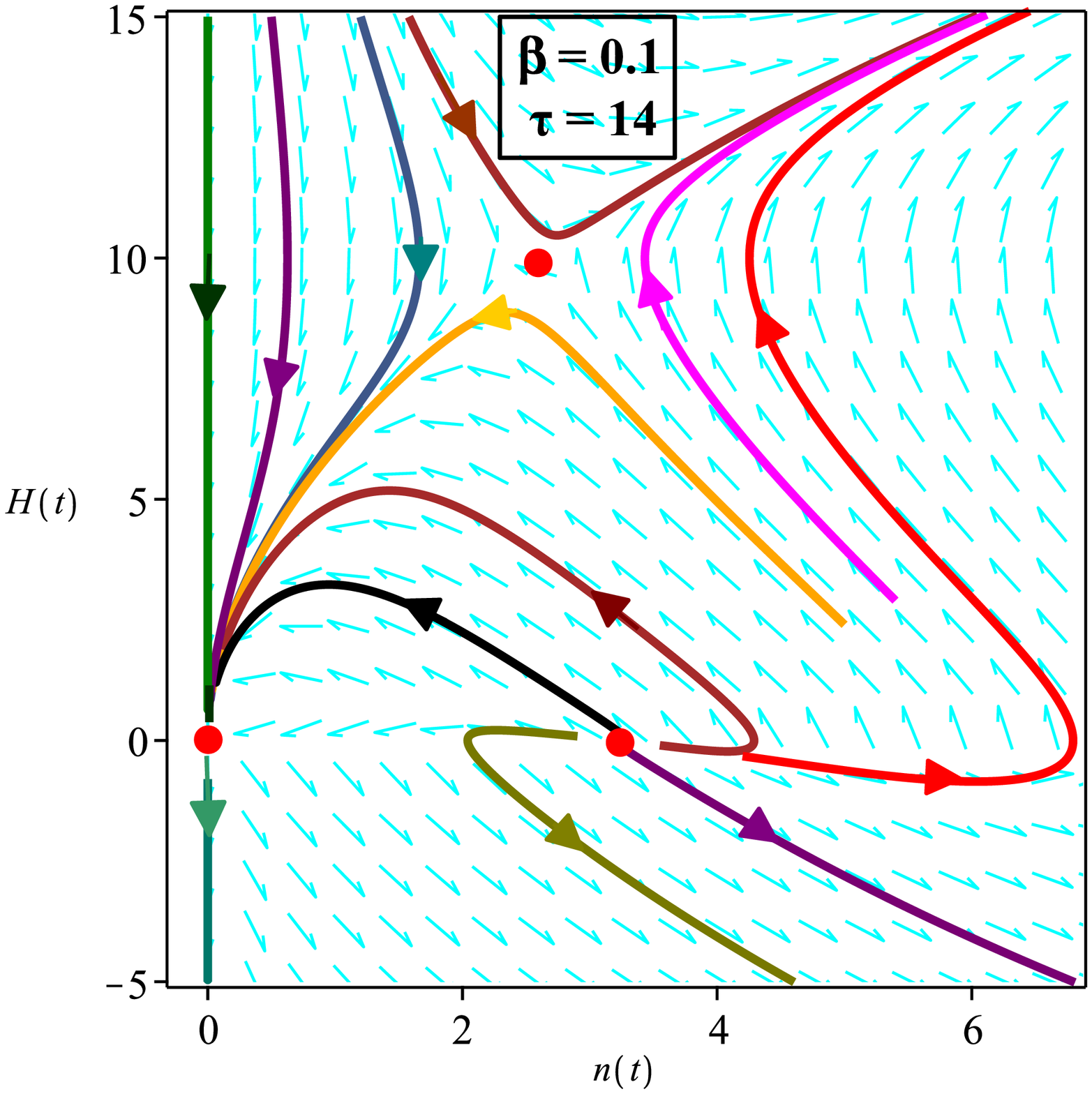}}
\,
\subfloat[\scriptsize Continuation of Figure 6a: at the saddle $(n^* = 80.4, H^* = 0)$, one has $\rho^* < 0$. The critical point $(n^{**} = 45.9,
H^{**} = 10)$ is a stable node.]
{\label{F6a2}\includegraphics[height=4.3cm,width=0.32\textwidth]{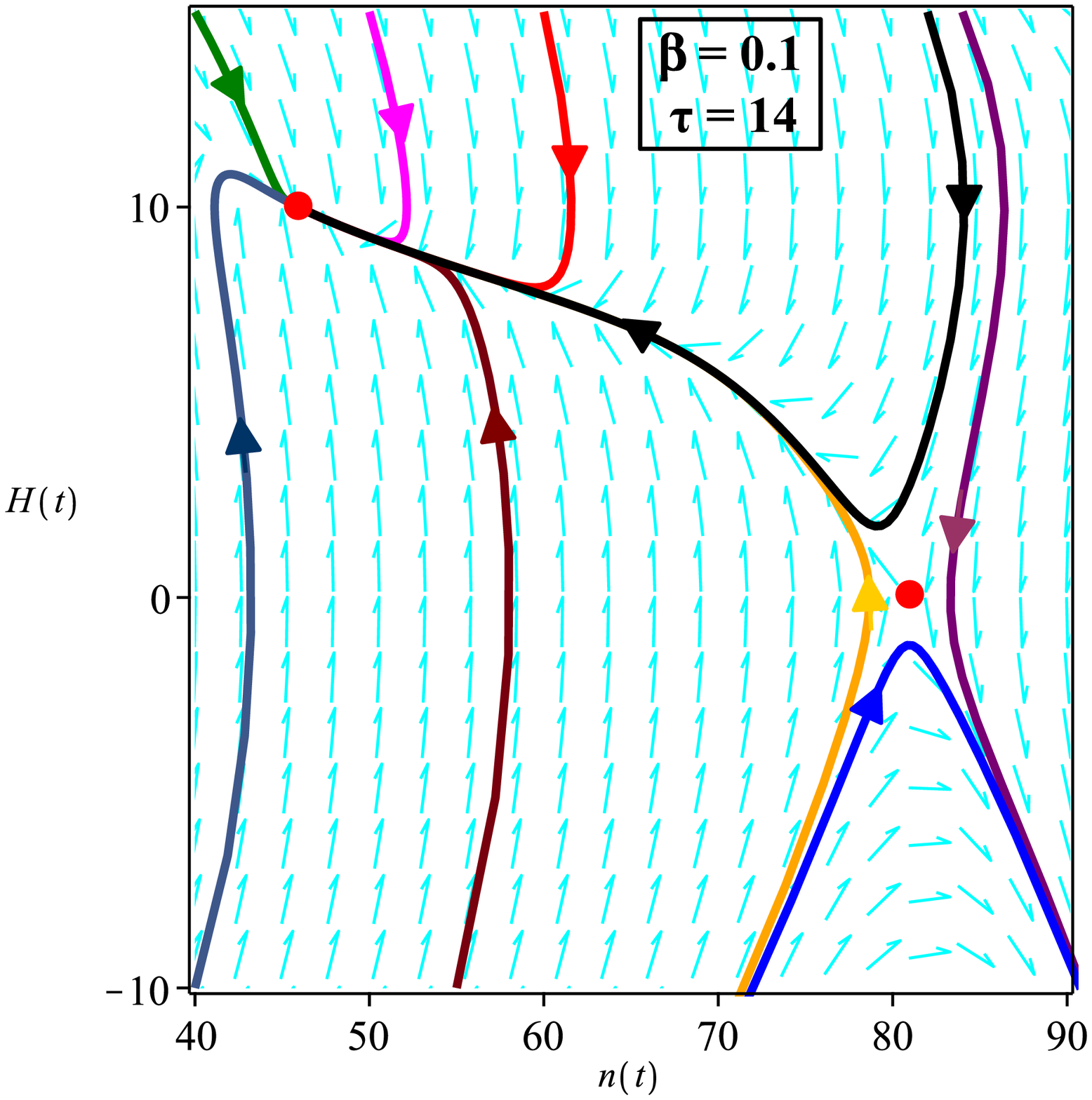}}
\,
\subfloat[\scriptsize Continuation of Figures 6a and 6b: the critical point $(n^{**} = 109.6, H^{**} = 10)$ is a saddle.]
{\label{F6a3}\includegraphics[height=4.3cm,width=0.32\textwidth]{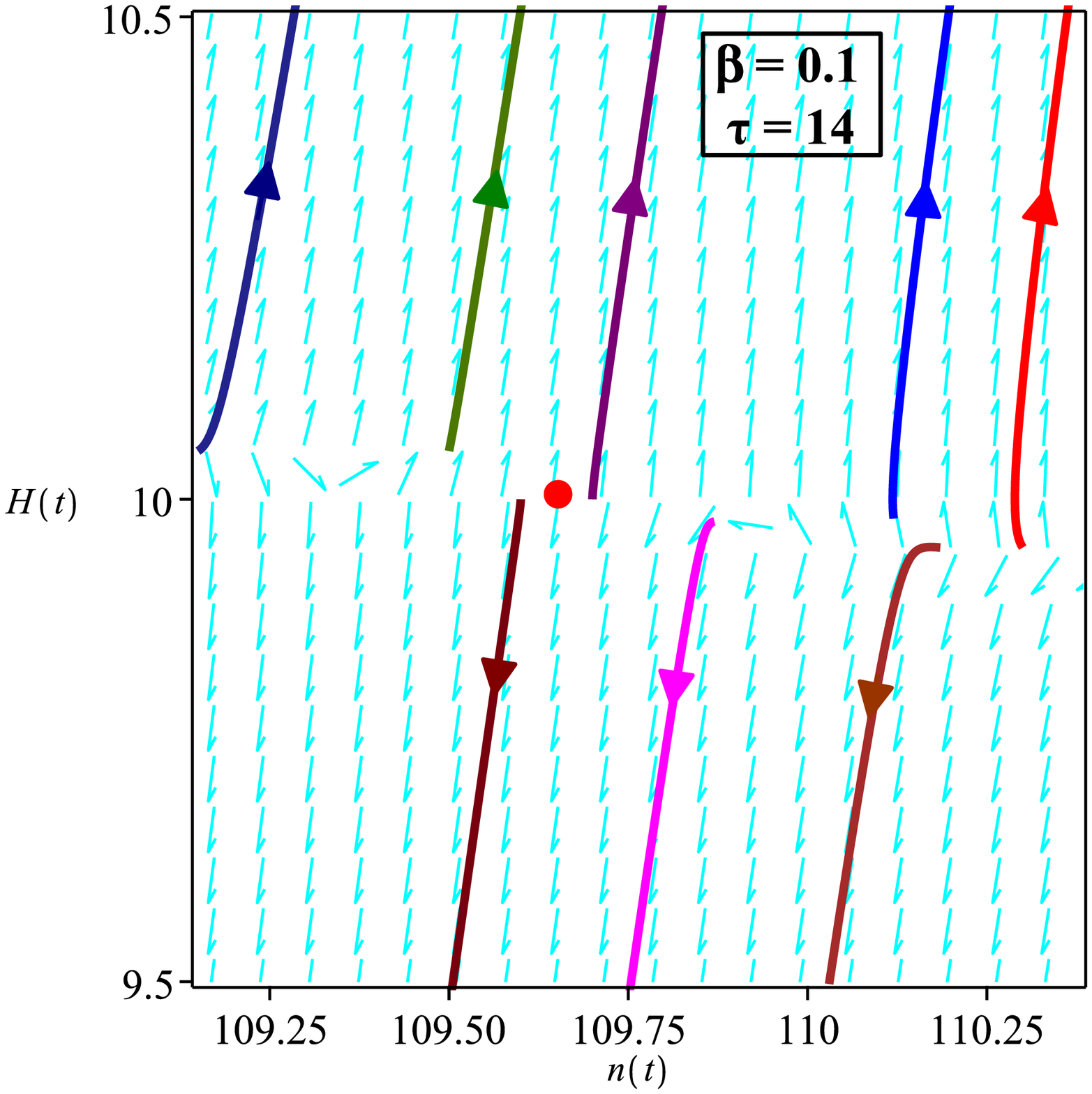}}
\caption{\footnotesize{Parts (a), (b), and (c) --- the case of $\alpha = 2$ with $\beta = 0.1$. As $\beta > \sqrt{8B}/m_0 = 0.0894$, the eigenvalues
$\lambda_{1,2}^*$ are real for all $n^*$ --- they are both positive for $0 < n^* <(\sqrt{3/2}-1)/A = 22.47$ (with the corresponding critical points
being unstable nodes, see Figure 6a and 6d) and positive and negative for $n^* > (\sqrt{3/2} - 1)/A = 22.47$ (with the corresponding critical points
being saddles, see Figure 6b, 6e). In relation to the eigenvalues $\lambda_{1,2}^{**}$, one has $n_1^{**} = 5.83$ and $n_2^{**} = 72.02$. Critical
points with $0 < n^{**} < n_1^{**} = 5.83$ are with real eigenvalues with opposite signs (saddles, see Figures 6a and 6d), those with $n^*$ between
$n_1^{**} = 5.83$ and $n_2^{**} = 72.02$ are with real and negative eigenvalues (stable nodes, see Figure 6b), and critical points with $n^*$ above
$n_2^{**} = 72.02$ are with real eigenvalues with opposite signs (saddles, see Figure 6c). }}
\label{Figure6_1}
\end{figure}

\addtocounter{figure}{-1}
\addtocounter{subfigure}{+3}

\begin{figure}[!ht]
\centering
\subfloat[\scriptsize There are four critical points when $\alpha = 2$, $\beta = 0.1$ and $\tau = 18$: the origin, the other two intersections of
$T^*(n^*)$ with $T(n^*)$, namely the unstable critical point $(n^* = 9.1, H^* = 0)$ (shown here), and a saddle at $(n^* = 46.5, H^* = 0)$, shown on
Figure 6e, and also the single intersection point of $T^{**}(n^{**})$ with $T(n^{**})$, namely the saddle at $(n^{**}=2.4, H^{**}=10)$.]
{\label{F6b1}\includegraphics[height=4.3cm,width=0.32\textwidth]{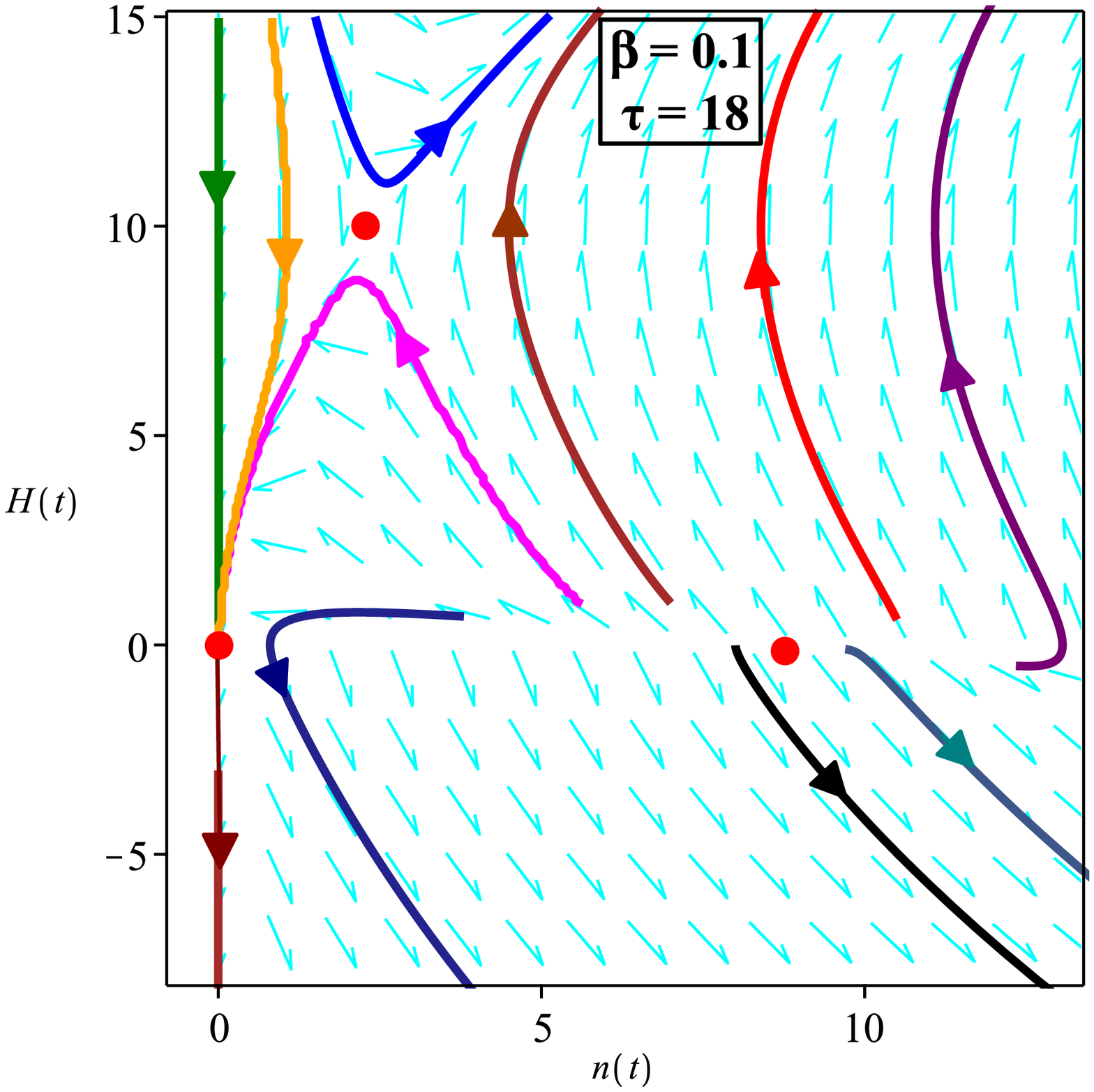}}
\,
\subfloat[\scriptsize Continuation of Figure 6d: the critical point at $(n^* = 46.5, H^* = 0)$ is a saddle. ]
{\label{F6b2}\includegraphics[height=4.3cm,width=0.32\textwidth]{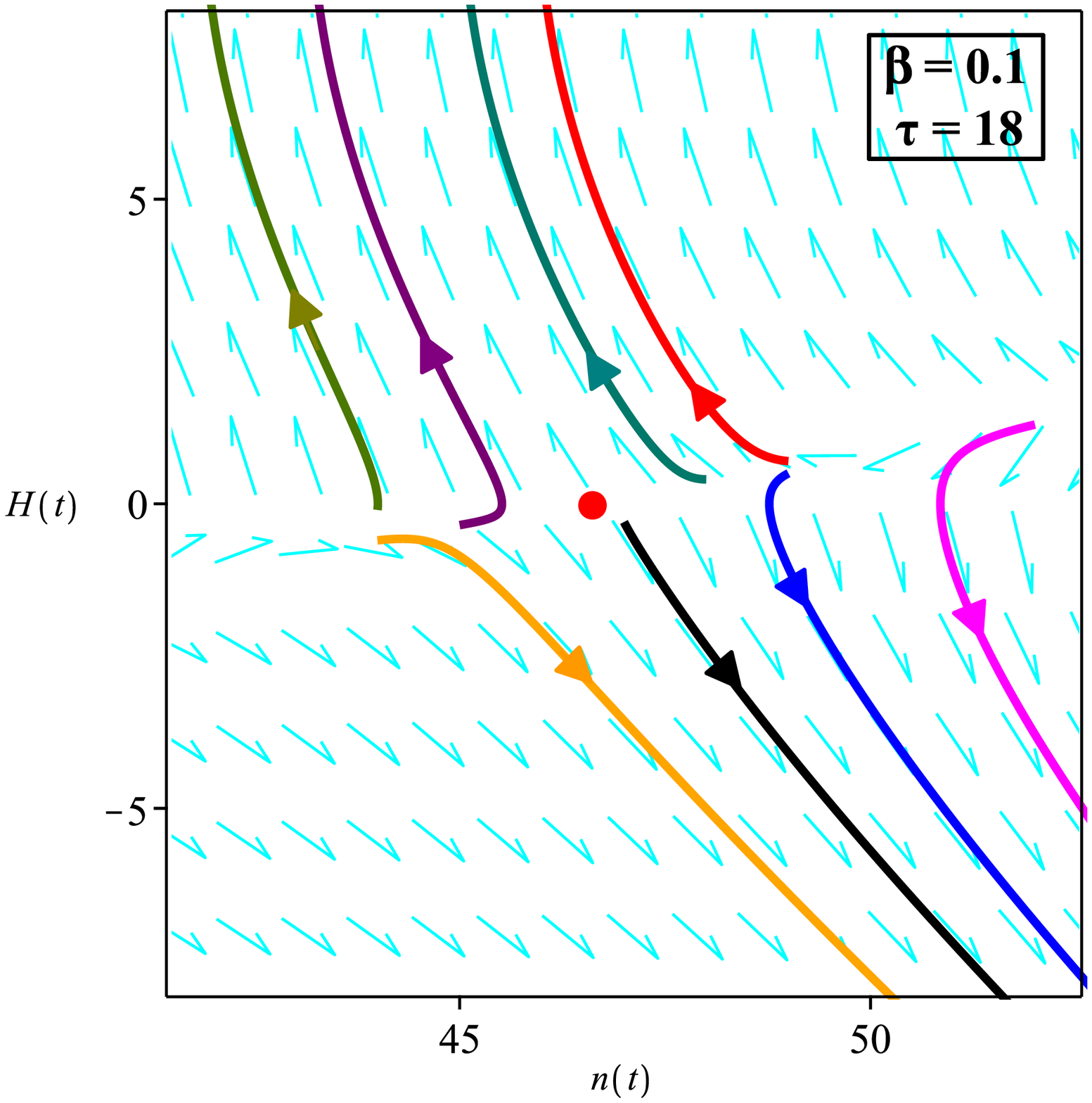}}
\,
\subfloat[\scriptsize There are two critical points when $\alpha = 2$, $\beta = 0.1$ and $\tau = 22$: the origin, which is the only intersection of
$T^*(n^*)$ with $T(n^*)$, and the single intersection point of $T^{**}(n^{**})$ with $T(n^{**})$ --- the saddle at $(n^{**} = 2.22, H^{**} = 10)$. ]
{\label{F6c}\includegraphics[height=4.3cm,width=0.32\textwidth]{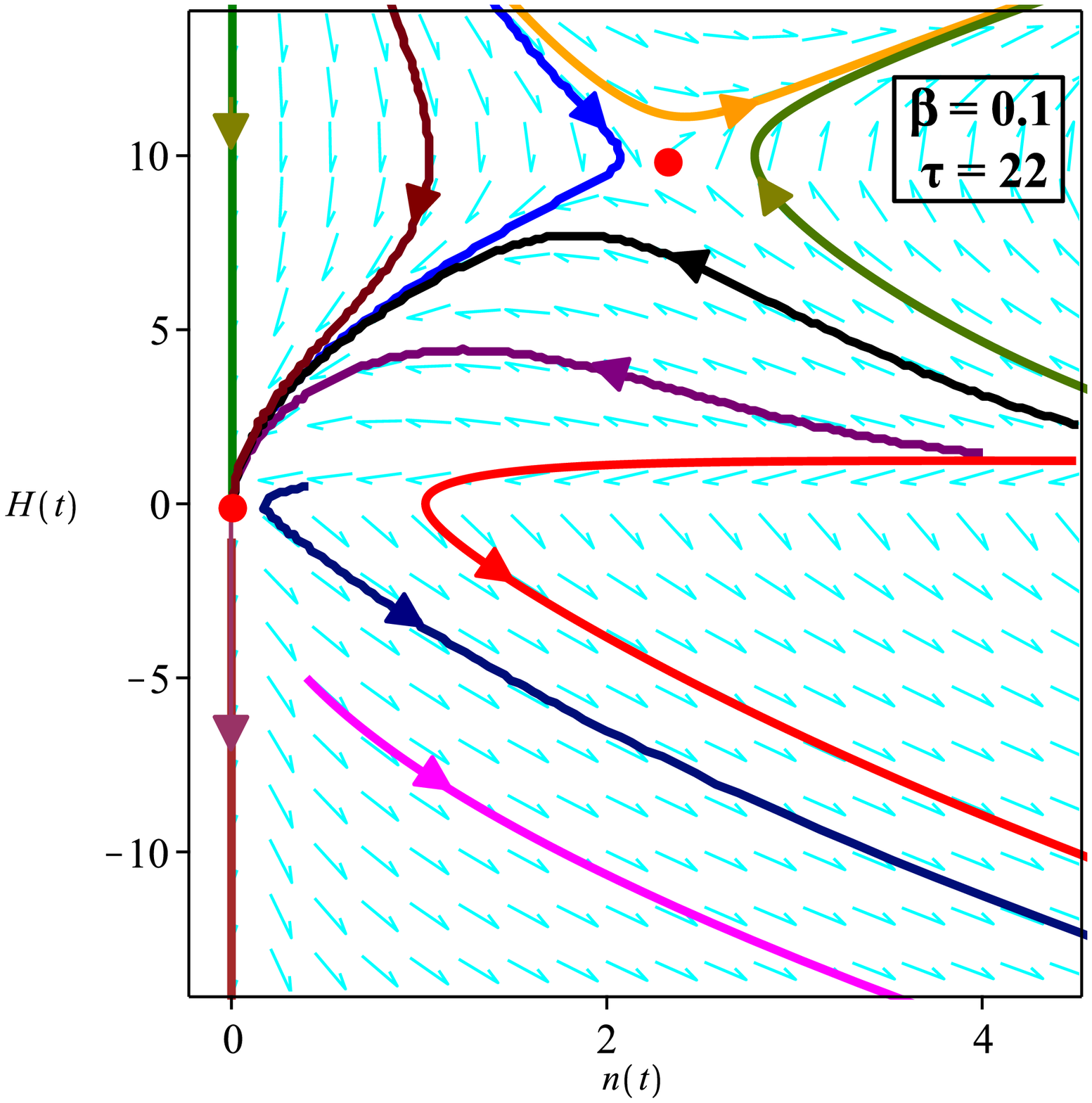}}
\caption{\footnotesize{Parts (d), (e), and (f) --- the case of $\alpha = 2$ with $\beta = 0.1$.  }}
\label{Figure6_2}
\end{figure}

\begin{figure}[!ht]
\centering
\subfloat[\scriptsize Here $\beta = 0.1 > \beta_0 = (12B/m_0^2)^{3/2} = 0.0013$. The positive roots of equation (\ref{nhat}) for $\hat{n}^{**}$ are $0.34$ and $73.52$. The respective solutions of equation (\ref{tauhat}) for $\hat{\tau}(\beta)$ are $-75.98$ and $14.78$. The first one is negative and leading to negative temperature and thus should be disregarded, that is, one should take $\hat{\tau}(\beta) = 14.78$. This corresponds to  $\hat{n}^{**} = 73.52$. On this diagram, $\tau = 12$ is taken and this is smaller than $\hat{\tau}(\beta) = 14.78$. Therefore, there are three intersection points of the curves $T^{**}(n^{**})$ and $T(n^{**})$ and thus three critical points of the type $(n^{**}, H^{**} = \beta^{1/(1-\alpha)})$, where $H^{**} = 2.15$ and $n^{**}$ is given by: $\hat{\nu}_0^{**} = 0.13$ (the saddle shown here), $\hat{\nu}_1^{**} = 33.09$, and $\hat{\nu}_2^{**} = 153.37$. The function $T^{**}(n^{**})$ is negative between $\nu_1^{**} = 0.14$ and $\nu_2^{**} = 9.86$. The intersection points of the curves $T^{**}(n^{**})$ and $Q(n^{**})$ are $\sigma_1^{**} = 0.34$ and $\sigma_2^{**} = 73.52$ (see also Figure 3) and between these two points, $Q(n^{**})$ is greater than $T^{**}(n^{**})$ and the critical points there are stable. But, as there can be no critical points of type $(n^{**}, H^{**} = \beta^{1/(1-\alpha)})$ when $T^{**}(n^{**}) < 0$, then all points between $\nu_2^{**} = 9.86$ and $\sigma_2^{**} = 73.52$ are stable (like the one at $n^{**} = \hat{\nu}_1^{**} = 33.09$ --- not shown). All others (like the one at $n^{**} = \hat{\nu}_0^{**} = 0.13$, shown, and the one at $n^{**} = \hat{\nu}_1^{**} = 153.37$, not shown) are saddles. The critical points of the type $(n^{*}, H^{*} = 0)$ are at $n_1^* = 1.90$ (shown here) and $n_2^* = 99.15$ (not shown). The first one, $n_1^* = 1.90$, is to the left of $\widetilde{n}^* = (\sqrt{3/2} - 1)/A = 22.48$ where the eigenvalues are purely imaginary and, as seen by the centre manifold theory, the trajectories are unstable spirals. The second one, $n_2^* = 99.15$, is to the right of $\widetilde{n}^* = 22.48$ where the eigenvalues are both real and with opposite signs, thus this critical point is a saddle.]
{\label{F7a}\includegraphics[height=5.8cm, width=0.48\textwidth]{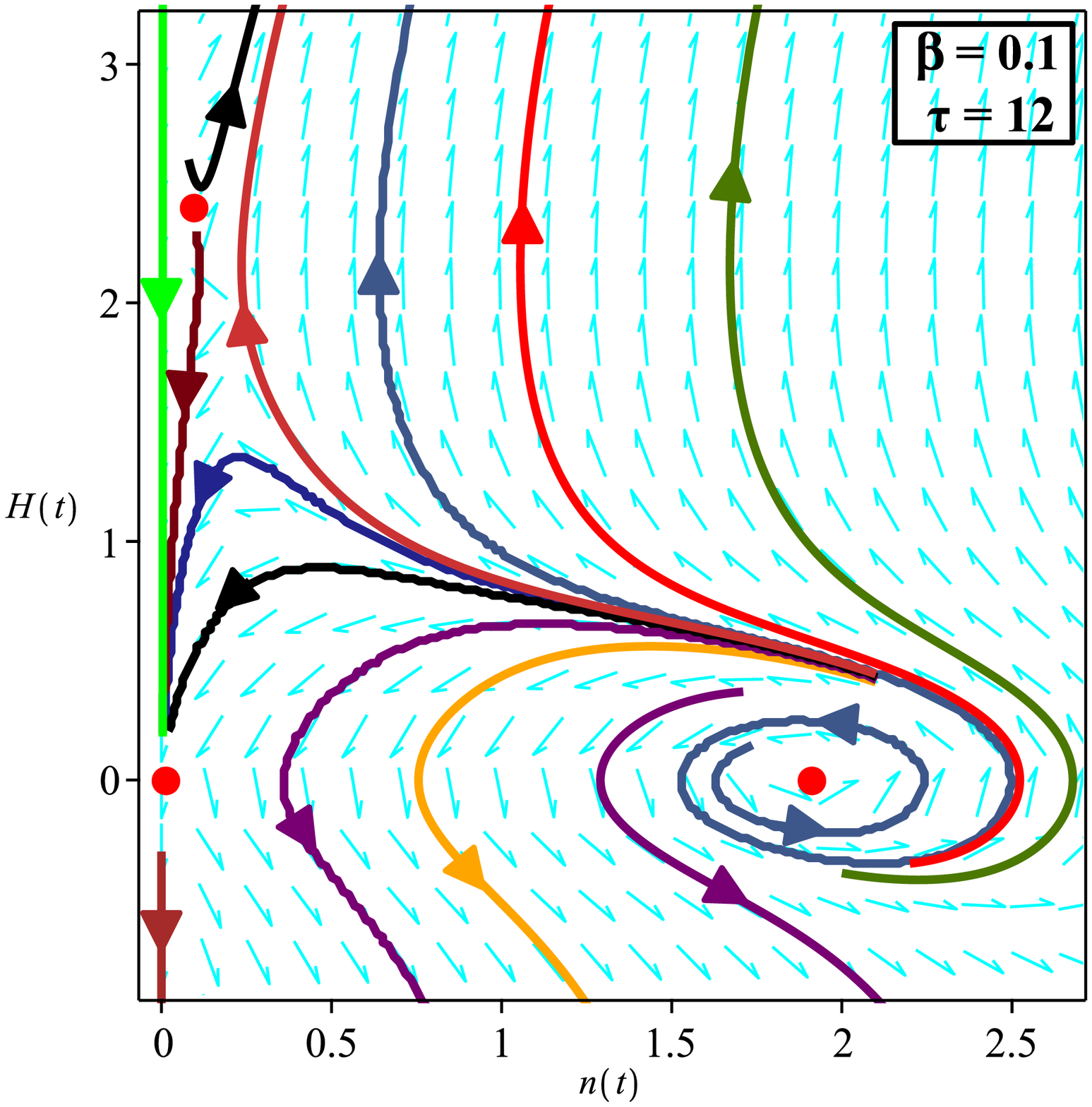}}
\quad
\subfloat[\scriptsize Here $\beta = 0.01 > \beta_0 = (12B/m_0^2)^{3/2} = 0.0013$. The positive roots of equation (\ref{nhat}) for $\hat{n}^{**}$ are $1.49$ and $73.26$. The respective solutions of equation (\ref{tauhat}) for $\hat{\tau}(\beta)$ are $-21.11$ and $14.80$. The first one is again negative and leading to negative temperature and thus should be disregarded, that is, one should take $\hat{\tau}(\beta) = 14.80$. This corresponds to  $\hat{n}^{**} = 73.26$. On this diagram, $\tau = 19.8$ is taken and this is greater than $\hat{\tau}(\beta) = 14.80$. Therefore, there is only one intersection point of the curves $T^{**}(n^{**})$ and $T(n^{**})$ and thus, there is just one critical point of the type $(n^{**}, H^{**} = \beta^{1/(1-\alpha)})$, where $H^{**} = 4.64$ and $n^{**} = \hat{\nu}_0^{**} = 0.56$. The function $T^{**}(n^{**})$ is negative between $\nu_1^{**} = 0.69$ and $\nu_2^{**} = 9.31$. The intersection points of the curves $T^{**}(n^{**})$ and $Q(n^{**})$ are $\sigma_1^{**} = 1.49$ and $\sigma_2^{**} = 73.26$ (see also Figure 3) and between these two points, $Q(n^{**})$ is greater than $T^{**}(n^{**})$. At point $n^{**} = \hat{\nu}_0^{**} = 0.56$, one has $T^{**}(n^{**}) > 0$, but $Q(n^{**}) < T^{**}(n^{**})$. Thus, the only critical point of type $(n^{**}, H^{**} = \beta^{1/(1-\alpha)})$ is not stable --- it is a saddle.
The critical points of the type $(n^{*}, H^{*} = 0)$ are at $n_1^* = 20.11$ and $n_2^* = 25.04$. They are shown on Figure 7c. ]
{\label{F7b}\includegraphics[height=5.8cm,width=0.48\textwidth]{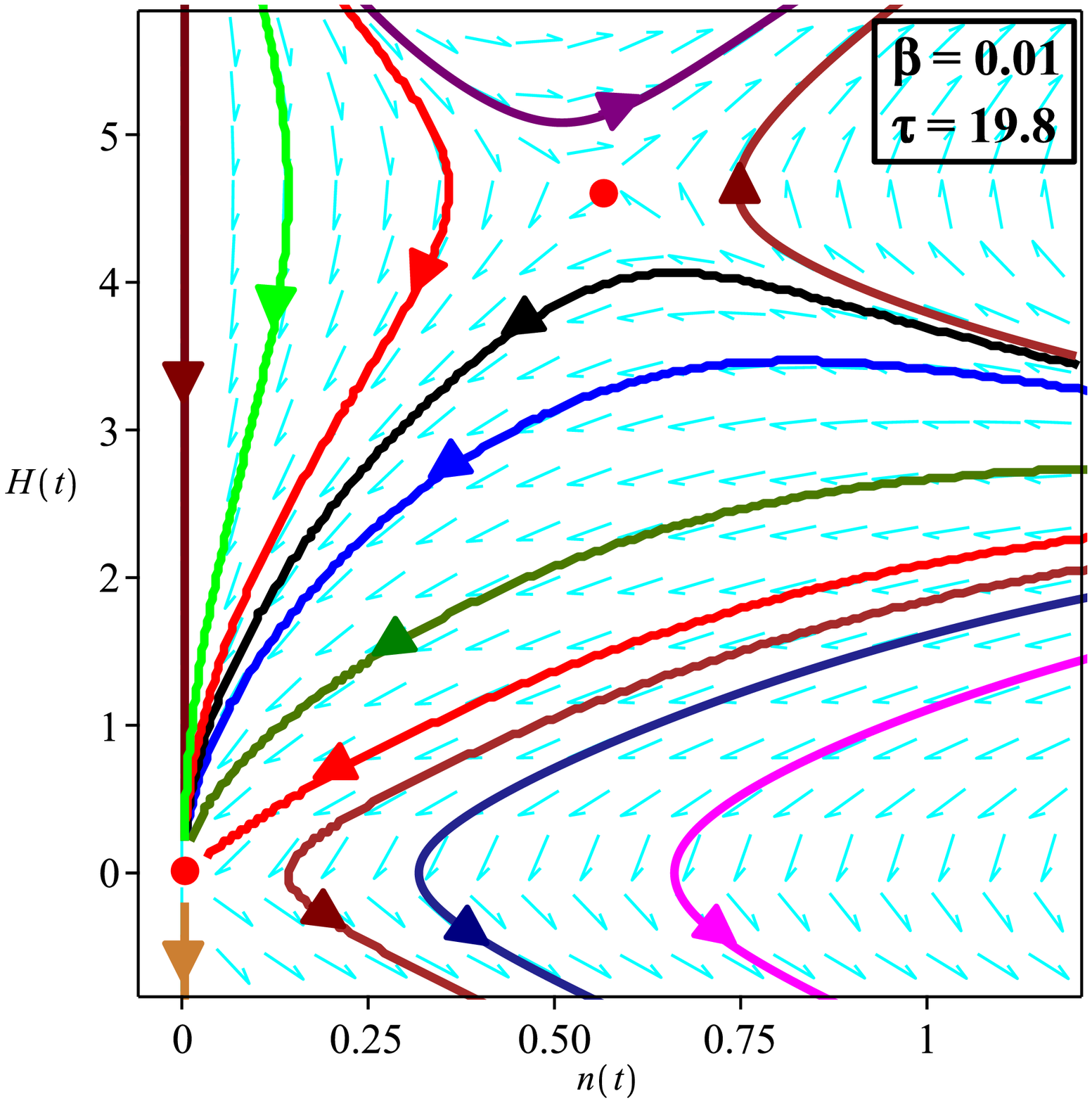}}
\caption{\footnotesize{Parts (a) and (b): The case of $\alpha = 4$ --- some representative cases.}}
\label{Figure7_1}
\end{figure}

\addtocounter{figure}{-1}
\addtocounter{subfigure}{+2}

\begin{figure}[!ht]
\centering
\subfloat[\scriptsize Continuation of Figure 7b. In addition to the saddle at $(n^{**}, H^{**} = \beta^{1/(1-\alpha)})$, where $H^{**} = 4.64$ and $n^{**} = \hat{\nu}_0^{**} = 0.56$, there are two critical points of the type $(n^{*}, H^{*} = 0)$: at $n_1^* = 20.11$ and at $n_2^* = 25.04$. The first of these, $n_1^* = 20.11$, is to the left of $\widetilde{n}^* = (\sqrt{3/2} - 1)/A = 22.48$ where the eigenvalues are purely imaginary and, as seen by the centre manifold theory, the trajectories are unstable spirals. The second one, $n_2^* = 25.04$, is to the right of $\widetilde{n}^* = 22.48$ where the eigenvalues are both real and with opposite signs, thus this critical point is a saddle. Such ``dipole" of unstable spirals and a saddle is always a present feature when $\tau < \widetilde{\tau} = 19.84$.]
{\label{F7d}\includegraphics[height=5.8cm, width=0.48\textwidth]{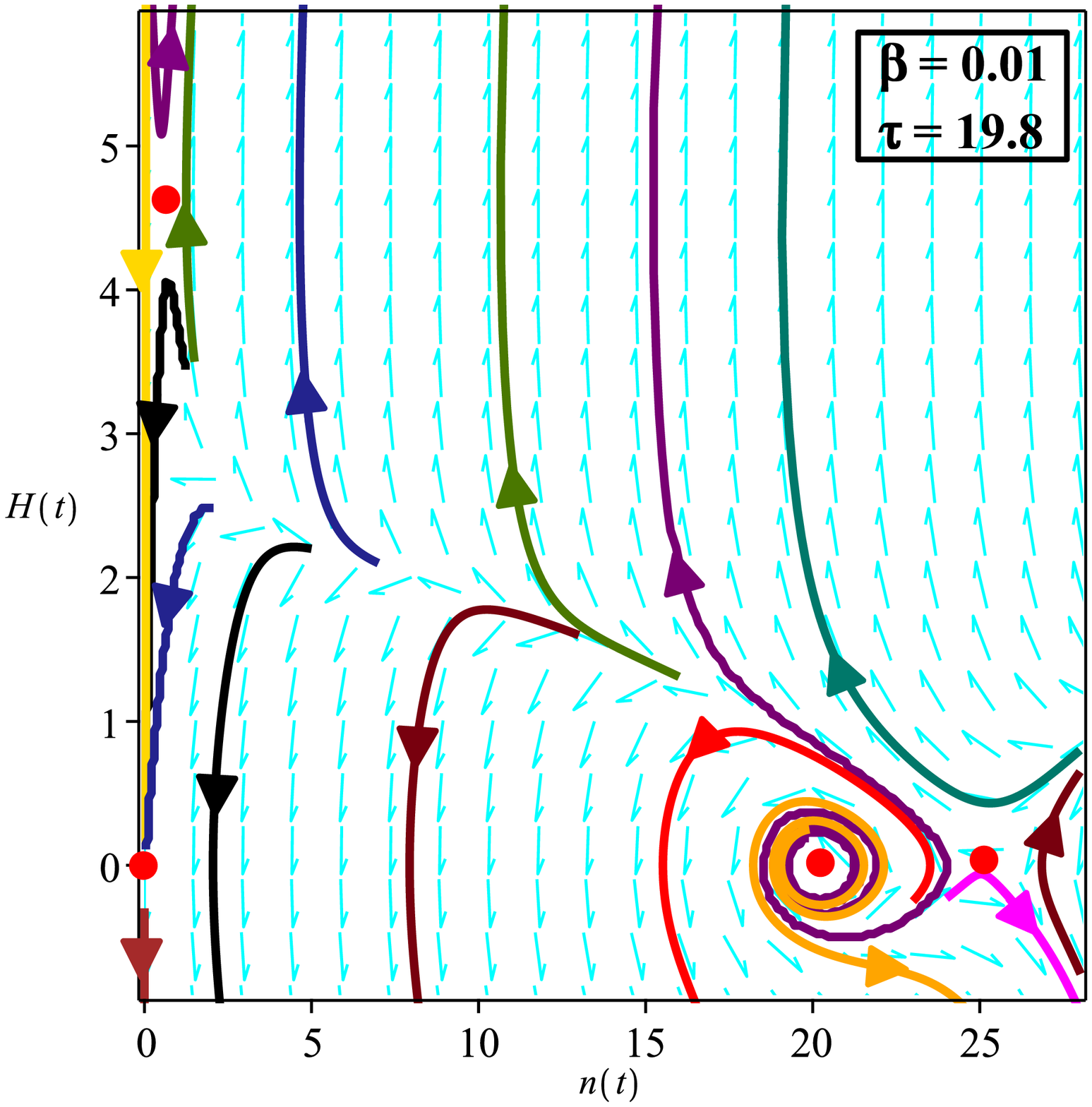}}
\quad
\subfloat[\scriptsize  Here $\beta = 0.05 > \beta_0 = (12B/m_0^2)^{3/2} = 0.0013$. The positive roots of equation (\ref{nhat}) for $\hat{n}^{**}$ are $0.54$ and $73.48$. The respective solutions of equation (\ref{tauhat}) for $\hat{\tau}(\beta)$ are $-53.76$ and $14.79$. The first one is negative and leading to negative temperature and thus should be disregarded, that is, one should take $\hat{\tau}(\beta) = 14.79$. This corresponds to  $\hat{n}^{**} = 73.48$. On this diagram, $\tau = 12$ is taken and this is smaller than $\hat{\tau}(\beta) = 14.79$. Therefore, there are three intersection points of the curves $T^{**}(n^{**})$ and $T(n^{**})$ and thus three critical points of the type $(n^{**}, H^{**} = \beta^{1/(1-\alpha)})$ with $n^{**}$ given by: $\hat{\nu}_0^{**} = 0.21$ (a saddle, not shown here), $\hat{\nu}_1^{**} = 48.71$ (the stable node shown here), and $\hat{\nu}_2^{**} = 108.68$ (a saddle, not shown here). The function $T^{**}(n^{**})$ is negative between $\nu_1^{**} = 0.23$ and $\nu_2^{**} = 9.77$. The intersection points of the curves $T^{**}(n^{**})$ and $Q(n^{**})$ are $\sigma_1^{**} = 0.54$ and $\sigma_2^{**} = 73.48$ (see also Figure 3) and between these two points, $Q(n^{**})$ is greater than $T^{**}(n^{**})$ and the critical points there are stable. But, as there can be no critical points of type $(n^{**}, H^{**} = \beta^{1/(1-\alpha)})$ when $T^{**}(n^{**}) < 0$, then all points between $\nu_2^{**} = 9.77$ and $\sigma_2^{**} = 73.48$ are stable --- including the one on the diagram at $n^{**} =  \hat{\nu}_1^{**} = 48.71$. The other two (not shown) are saddles. The critical points of the type $(n^{*}, H^{*} = 0)$ are at $n_1^* = 3.22$ and at $n_2^* = 80.38$. None of them are shown here. The first one, $n_1^* = 3.22$, is to the left of $\widetilde{n}^* = (\sqrt{3/2} - 1)/A = 22.48$ where the eigenvalues are purely imaginary and, as seen by the centre manifold theory, the trajectories are unstable spirals. The second one, $n_2^* = 80.38$, is to the right of $\widetilde{n}^* = 22.48$ where the eigenvalues are both real and with opposite signs, thus this critical point is a saddle. One has again a ``dipole" of unstable spirals and a saddle.]
{\label{F7e}\includegraphics[height=5.8cm,width=0.48\textwidth]{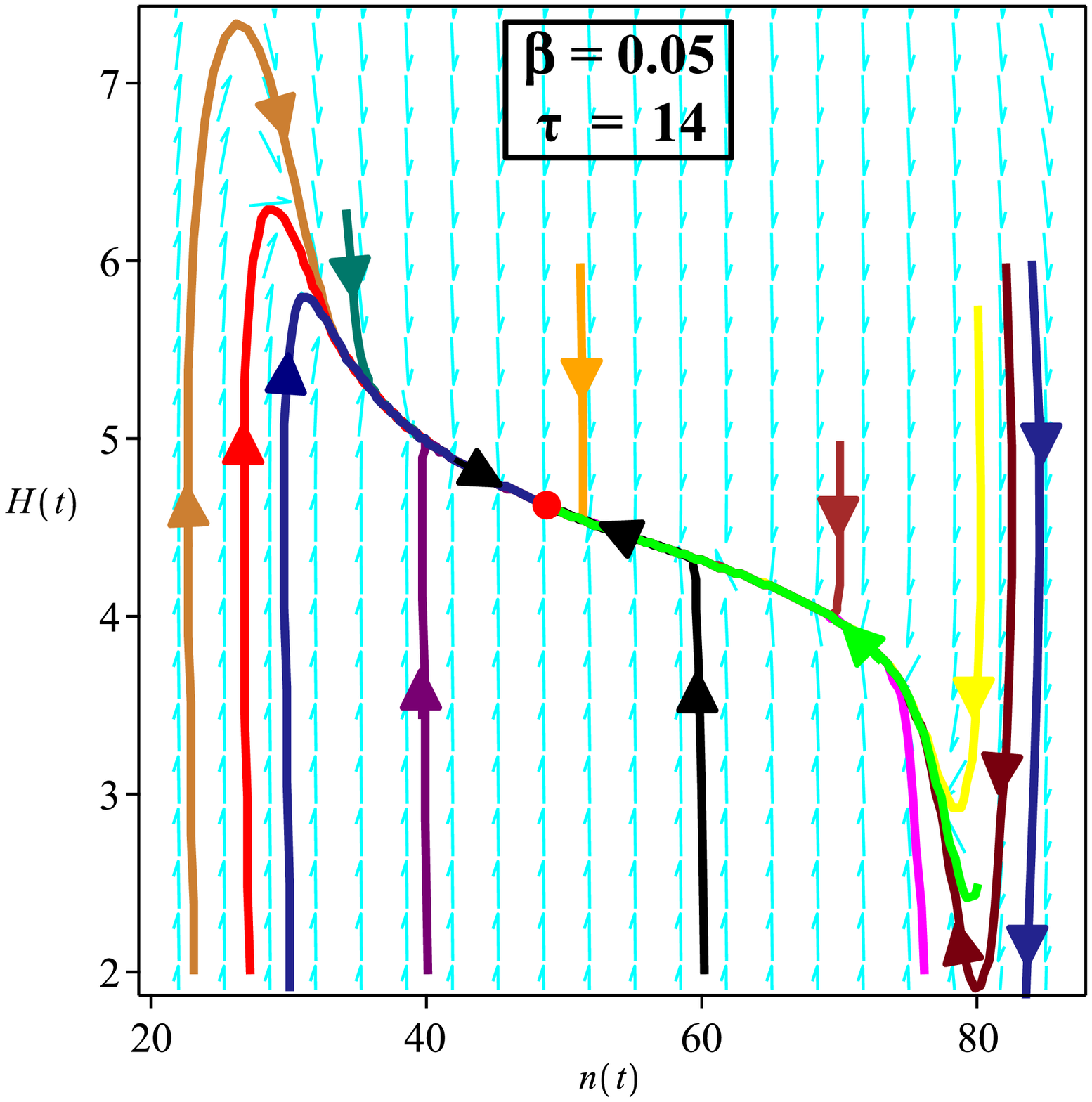}}
\caption{\footnotesize{Parts (c) and (d): The case of $\alpha = 4$ --- some representative cases.}}
\label{Figure7_2}
\end{figure}

\section{Conclusions}

A cosmological model with two matter components --- dust and gas with van der Waals equation of state has been examined. In addition, the model includes a particle production term, proportional to a constant power, $\alpha$, of the Hubble parameter $H$. Models with $\alpha=2$ and $\alpha = 4$ are studied in detail. However, the presented analysis can easily be extended to an arbitrary integer $\alpha$ (the special case of $\alpha=1$ deserves a special attention and will be provided elsewhere). \\
The time-evolution of the model is given by a nonlinear dynamical system of three equations: for the particle number density $n$, the Hubble parameter $H$ and the temperature $T$. This system admits a global first integral, which explicitly gives $T$ as a function of $n$ and one of the van der Waals gas parameters. Hence, the system is reduced to a two-component one: in the two dimensional $n$--$H$ phase space. The system exhibits a complex behavior which is influenced by the presence of the several model parameters. This behaviour is examined in detail using the phase-plane analysis for all possible parameter choices. The two second integrals of the system are represented by curves which separate the phase space into domains which can not be crossed by the trajectories. The full classification of the critical points is presented in the two provided tables.
It is shown that the critical points can not be reached in a finite time (the stable critical points can only be reached for $t\to \infty$, the unstable critical points can be reached only for $t \to -\infty.$). The critical points provide important information about the large-time behaviour of the system. This includes both the distant future ($t \to \infty$) or the distant past ($t \to -\infty$). For example, considering trajectories which end at the origin, i.e. $(n,H) \to (0,0)$ as $t \to \infty$, from (\ref{nn}) one has $\dot{n}=-3nH$ asymptotically when $\alpha \ge 2$ and taking into account (\ref{ce-d}), it follows that $d \rho_d / dn = \rho_d / n,$ or $\rho_d = C n $ for some constant $C$. Then
\b
\frac{\rho_d}{\rho}=\frac{Cn}{ n [m_0 + \frac{3}{2} \, \tau \, n^{\frac{2}{3}} \, e^{\frac{2 A n}{3}} - B n ]} \to \frac{C}{m_0} = \mathrm{const}
\e 
when $(n,H) \to (0,0).$  Therefore, the ratio between the two fractions approaches a constant. \\
In the case of high particle creation $n \to \infty$ and  $H \to \infty$ (this can be viewed as a critical point at infinity), in the distant past or future, i.e. when $t \to \pm \infty$, the asymptotic equations are
\b
\dot{n} & = & 3 \, \beta \, n \,  H^{\alpha}, \\ 
\dot{H} & = & \frac{1}{2} \, \beta A \, \tau \, n^{\frac{8}{3}} \, e^{\frac{2An}{3}} \, H^{\alpha -1},
\e
giving $H^2 = (1/2) \, \tau \, n ^{5/3} \, e^{2An/3} \, + \, $ lower-order terms. Substituting this asymptotic form of $H^2$ into the Friedmann equation (\ref{h2}) yields:
\b
\frac{1}{3} \left(1 + \frac{\rho_d}{\rho}\right)=\frac{\frac{1}{2} \, \tau \, n ^{\frac{5}{3}} e^{\frac{2An}{3}} + \ldots }{n [m_0 + 
\frac{3}{2} \, \tau \, n ^{\frac{5}{3}} e^{\frac{2An}{3}} - B n]} \to \frac{1}{3} \quad \mbox{ when $(n, H) \to (\infty, \infty)$.}
\e
In other words $\rho_d /\rho \to 0,$ which means that in this case the dust component becomes negligible and all trajectories are drawn in the neighbourhood of the separatrix $3H^2 = \rho$ as  $t \to \pm \infty$. \\
Finally, sets of initial values can be identified for which the corresponding trajectories exhibit inflationary behavior.

\begin{center}
\begin{sidewaystable}[ht]
\begin{tabular}{|l|c|c|c|c|c|l|c|c|}
\hline
\multicolumn{1}{|c|}{\begin{tabular}[c]{@{}c@{}}Critical\\ Points\end{tabular}} & \multirow{4}{*}{Parameters}            & \multicolumn{5}{c|}{$\alpha = 2$}                                                                                                                                                                                                                                                                                                                                                                                                                                                                                                                               & \multicolumn{2}{c|}{$\alpha = 4$}                                                                                                                                                                                                                                 \\ \cline{1-1} \cline{3-9}
\multirow{5}{*}{$(n^* \ne 0, H^*=0)$}                                           &                                        & \multicolumn{5}{c|}{$\beta$}                                                                                                                                                                                                                                                                                                                                                                                                                                                                                                                                    & \multirow{3}{*}{$n^*<\widetilde{n}^*$}                                                                                                                                      & \multirow{3}{*}{$n^*>\widetilde{n}^*$}                                              \\ \cline{3-7}
                                                                                &                                        & \multicolumn{3}{c|}{$\beta<\frac{\sqrt{8B}}{m_0}$}                                                                                                                                                                                                                                                                                                                   & \multicolumn{2}{c|}{$\beta>\frac{\sqrt{8B}}{m_0}$}                                                                                                                         &                                                                                                                                                                             &                                                                                     \\ \cline{3-7}
                                                                                &                                        & $0<n^*<N_0^*$                                                                                                            & $N_0^*<n^*<\widetilde{n}^*$                                                                                                           & $n^*>\widetilde{n}^*$                                                                                    & \multicolumn{1}{l|}{$n^*<\widetilde{n}^*$}                                        & $n^*>\widetilde{n}^*$                                                                         &                                                                                                                                                                             &                                                                                     \\ \cline{2-9}
                                                                                & $\tau < \widetilde{\tau}$      & \multicolumn{2}{c|}{\begin{tabular}[c]{@{}c@{}}Always \\ existing: \\ \\ \\ either \\ \\ \\ unstable spiral \\ in $0<n^*<N_0^*$ \\ --- Fig. 2a, 2b, \\ 4a, 5a, 5d,\\ \\ or\\  \\ unstable node \\ in $N_0^*<n^*<\widetilde{n}^*$,\\ --- Fig 2a, 2b\end{tabular}} & \begin{tabular}[c]{@{}c@{}}Always \\ existing \\ saddle,\\ Fig. 2a, 2b, \\ 4c,5c, \\ 5e, 6e\end{tabular} & \begin{tabular}[c]{@{}c@{}}Unstable \\ node,\\ Fig. 2a, 2c,\\ 6a, 6d\end{tabular} & \multicolumn{1}{c|}{\begin{tabular}[c]{@{}c@{}}Saddle,\\ Fig. 2a, 2c, \\ 6b, 6e\end{tabular}} & \begin{tabular}[c]{@{}c@{}}Always \\ existing\\ unstable \\ spiral\\ with purely\\ imaginary \\ eigenvalues\\ (centre \\ manifold\\ theory) \\ --- Fig. 7a, 7c\end{tabular} & \begin{tabular}[c]{@{}c@{}}Always \\ existing \\ saddle \\ --- Fig. 7c\end{tabular} \\ \cline{2-9}
                                                                                & $\tau > \widetilde{\tau}$      & \multicolumn{5}{c|}{\begin{tabular}[c]{@{}c@{}}Does not exist, \\ Fig. 2a, 4c, 5f, 6f\end{tabular}}                                                                                                                                                                                                                                                                                                                                                                                                                                                             & \multicolumn{2}{c|}{Does not exist}                                                                                                                                                                                                                               \\ \hline
\multicolumn{1}{|c|}{\multirow{2}{*}{$(0,0)$}}                                  & \multicolumn{1}{l|}{\multirow{2}{*}{}} & \multicolumn{7}{c|}{Attracts trajectories from the upper half-plane $H > 0$}                                                                                                                                                                                                                                                                                                                                                                                                                                                                                                                                                                                                                                                                                                                                                        \\ \cline{3-9}
\multicolumn{1}{|c|}{}                                                          & \multicolumn{1}{l|}{}                  & \multicolumn{7}{c|}{Repels trajectories from the lower half-plane $H < 0$}                                                                                                                                                                                                                                                                                                                                                                                                                                                                                                                                                                                                                                                                                                                                                           \\ \hline
\end{tabular}
\end{sidewaystable}
\end{center}

\begin{center}
\begin{sidewaystable}[ht]
\begin{tabular}{|c|c|c|c|c|c|c|}
\hline
\multirow{2}{*}{Critical Points}                               & \multirow{2}{*}{Parameters}                 & \multirow{2}{*}{$T(n^{**}) = T^{**}(n^{**})$ at:} & \multicolumn{4}{c|}{$\alpha = 2$ and $\alpha = 4$}                                                                                                                                                                                                                                                                                                                                                                                                                      \\ \cline{4-7}
                                                               &                                             &                                                   & $\beta > \beta_0$                                                                                                   & $\beta = \beta_0$                                                                                                 & $\beta_Q \le \beta < \beta_0$                                                                                    & $\beta<\beta_Q$                                                                                            \\ \hline
\multirow{4}{*}{$(n^{**}, H^{**}=\beta^{\frac{1}{\alpha-1}})$} & $\tau > \hat{\tau}(\beta)$                  & $n^{**}= \hat{\nu}_0^{**}$                        & \begin{tabular}[c]{@{}c@{}}Saddle,\\ $\hat{\nu}_0^{**}<\nu_1^{**}$\\ Fig. 3a, 3b\end{tabular}                       & \begin{tabular}[c]{@{}c@{}}Saddle,\\ $\hat{\nu}_0^{**}<\nu_1^{**}$\\ Fig. 3a, 3b\end{tabular}                     & \begin{tabular}[c]{@{}c@{}}Saddle,\\ $\hat{\nu}_0^{**}<\nu_1^{**}$\\ Fig. 3a, 3b\end{tabular}                    & \begin{tabular}[c]{@{}c@{}}Saddle,\\ $\hat{\nu}_0^{**}<\nu_1^{**}$\\ Fig. 3a, 3b\end{tabular}              \\ \cline{2-7}
                                                               & \multirow{3}{*}{$\tau < \hat{\tau}(\beta)$} & $n^{**}=\hat{\nu}_0^{**}$                         & \begin{tabular}[c]{@{}c@{}}Saddle,\\ $\hat{\nu}_0^{**}<\nu_1^{**}$\\ Fig3b, 3c, \\ 7a, 7b, 7c\end{tabular}          & \begin{tabular}[c]{@{}c@{}}Saddle,\\ $\hat{\nu}_0^{**}<\nu_1^{**}$\\ Fig3b, 3c, \\ 7a, 7b, 7c\end{tabular}        & \begin{tabular}[c]{@{}c@{}}Saddle,\\ $\hat{\nu}_0^{**}<\nu_1^{**}$\\ Fig3b, 3c, \\ 7a, 7b, 7c\end{tabular}       & \begin{tabular}[c]{@{}c@{}}Saddle,\\ $\hat{\nu}_0^{**}<\nu_1^{**}$\\ Fig3b, 3c, \\ 7a, 7b, 7c\end{tabular} \\ \cline{3-7}
                                                               &                                             & $n^{**}=\hat{\nu}_1^{**}$                         & \begin{tabular}[c]{@{}c@{}}Stable, \\ $\nu_2^{**} < \hat{\nu}_1^{**}<\sigma_2^{**}$,\\ Fig. 3b, 3c, 7d\end{tabular} & \begin{tabular}[c]{@{}c@{}}Stable, \\ $\nu_0^{**} < \hat{\nu}_1^{**}<\chi_2^{**}$,\\ Fig. 3b, 3c, 7d\end{tabular} & \begin{tabular}[c]{@{}c@{}}Stable, \\ $\xi_1^{**} < \hat{\nu}_1^{**}<\xi_2^{**}$,\\ Fig. 3b, 3c, 7d\end{tabular} & \begin{tabular}[c]{@{}c@{}}Saddle\\ for all $n^{**}$\end{tabular}                                          \\ \cline{3-7}
                                                               &                                             & $n^{**}=\hat{\nu}_2^{**}$                         & \begin{tabular}[c]{@{}c@{}}Saddle, \\ $\hat{\nu}_2^{**}>\sigma_2^{**}$,\\ Fig. 3b, 3c\end{tabular}                  & \begin{tabular}[c]{@{}c@{}}Saddle, \\ $\hat{\nu}_2^{**}>\chi_2^{**}$,\\ Fig. 3b, 3c\end{tabular}                  & \begin{tabular}[c]{@{}c@{}}Saddle, \\ $\hat{\nu}_2^{**}>\xi_2^{**}$,\\ Fig. 3b, 3c\end{tabular}                  & \begin{tabular}[c]{@{}c@{}}Saddle\\ for all $n^{**}$\end{tabular}                                          \\ \hline
\end{tabular}
\end{sidewaystable}
\end{center}

\newpage

\begin{center}
{\bf \large Appendix} \vskip.3cm 
{\it Application of Centre-Manifold Theory to the Critical Points with Purely Imaginary Eigenvalues}
\end{center}

\n
To study the behaviour of the trajectories near the critical points $(n^*, H^* = 0)$ for which the eigenvalues are purely imaginary, namely, for
$n^* < (\sqrt{3/2} - 1)/A$, centre-manifold theory \cite{aaa} is applied.
Firstly, the dynamical system is expanded near
such $(n^*, H^* =0)$:
\b
(n - n^*)^{\bigdot} & = & - 3 n^* H - 3 (n - n^*) H (1 - \beta H^3) + 3 \beta n^* H^4, \\
\dot{H} & = & - \frac{3}{2} H^2 + \frac{1}{2} \beta \rho^* H^3 + \frac{Bn^* ( A^2 n^{*^2} + 2An^* - \frac{1}{2})}{An^*+1}(n - n^*) \nonumber \\
&& - \frac{1}{9} \, \frac{B( A^3 n^{*^3} + 9 A^2 n^{*^2}  + \frac{21}{2} An^* - 2)}{An^*+1}(n - n^*)^2 \nonumber \\
&& - \frac{1}{162} \, \frac{B( 4 A^4 n^{*^4} + 52  A^3 n^{*^3} + 150 A^2 n^{*^2}  + 70 An^* - 5)}{(An^*+1) n^*}(n - n^*)^3 + \ldots. \nonumber \\
\e
Introduce new dynamical variables via: $n - n^* = \theta x$ and $H = \mu y$. Introduce also $\omega = 3n^* \mu / \theta$. Taking $\theta = 1$ and $\mu
= (1/3) \sqrt{- B ( A^2 n^{*^2} + 2An^* - \frac{1}{2}) / (A n^* + 1)}$ [note that $\mu$ is real to the left of $\widetilde{n}^* = (\sqrt{3/2}-1)/A$
--- where the analysis applies]. This yields $\omega^2 = - B n^* ( A^2 n^{*^2} + 2An^* - \frac{1}{2}) / (A n^* + 1) > 0$. \\
The dynamical system can then be written as:
\b
\dot{x} & = & - \omega y + f(x, y), \\
\dot{y} & = & \omega x + g(x, y),
\e
where
\b
f(x, y) & = & - 3 \mu x y (1 - \beta \mu^3 y^3) + 3 \beta n^* \mu^4 y^4 ,\\
g(x, y) & = & - \frac{3}{2} \mu y^2 + \frac{1}{2} \mu^2 \beta \rho^* y^3 - \frac{1}{9 \mu} \, \frac{B( A^3 n^{*^3} + 9 A^2 n^{*^2}
+ \frac{21}{2} An^* - 2)}{An^*+1}x^2 \nonumber \\
&& - \frac{1}{162 \mu} \, \frac{B( 4 A^4 n^{*^4} + 52  A^3 n^{*^3} + 150 A^2 n^{*^2}  + 70 An^* - 5)}{(An^*+1) n^*}x^3 + \ldots.
\e
Then, at the critical point $(x = 0, y = 0)$, the stability parameter $a$ --- see (3.4.10) and (3.4.11) in \cite{aaa} --- is :
\b
a & = & \frac{1}{16} \, (f_{xxx} + f_{xyy} + g_{xxy} + g_{yyy}) \nonumber \\
& & + \,\, \frac{1}{16 \omega} \, [f_{xy} (f_{xx} + f_{yy}) - g_{xy} (g_{xx} + g_{yy}) - f_{xx} g_{xx} + f_{yy} g_{yy}] \nonumber \\
& = & \frac{3}{16} \mu^2 \beta \rho^* > 0,
\e
provided $\rho^* > 0$. \\
The energy density $\rho^*$ at the equilibrium points $(n^*, H^* = 0)$ is, as in the case of $\alpha = 2$, non-negative for $n^* < n_0^* = \bigl( 2
m_0 A + B   + \sqrt{4 m_0^2 A^2 + 20 m_0 A B  + B^2} \, \bigr) / (4AB)$. And, given that one always has  $n_0^* > \widetilde{n}^* = (\sqrt{3/2}-1)/A$,
then $\rho^*$ is positive in the entire region $0 < n^* < \widetilde{n}^* = (\sqrt{3/2}-1)/A$ --- where the eigenvalues are purely imaginary. This, in turn, means that $a$ is positive in that region and thus all critical points with purely imaginary eigenvalues are unstable --- the trajectories near them are unwinding spirals \cite{aaa} --- see Figures 7a and 7c.


\begin{thebibliography}{99}

\bibitem{engoliam} L. Parker, {\it Particle Creation in Expanding Universes}, Phys. Rev. Lett. {\bf 21}, 562 (1968); \newline
Ya.B. Zeldovich, {\it Particle Production in Cosmology}, JETP Lett. {\bf 12}, 307 (1970).

\bibitem{prigo} I. Prigogine, J. Geheniau, E. Gunzig, and P. Nardone, Gen. Relativ. Gravit. {\bf 21(8)}, 767 (1989).

\bibitem{equiva1} M.O. Calv\~ao, J.A.S. Lima, and I. Waga, {\it On the Thermodynamics of Matter Creation in Cosmology}, Phys. Lett. {\bf A 162(3)}, 
223-226 (1992).
    
\bibitem{equiva2} J.A.S. Lima and A.S.M. Germano, {\it On the Equivalence of Bulk Viscosity and Matter Creation}, Phys. Lett. {\bf A 170}, 373 (1992).

\bibitem{zimdahl} W. Zimdahl, {\it Cosmological Particle Production, Causal Thermodynamics, and Inflationary Expansion}, Phys. Rev. {\bf D 61}, 083511
(2000), arXiv:astro-ph/9910483.

\bibitem{nie2} R.I. Ivanov and E.M. Prodanov, {\it Integrable Cosmological Model with a Van Der Waals Gas and Matter Creation}, submitted for publication.

\bibitem{vilasi} G. Vilasi, {\it Hamiltonian Dynamics}, World Scientific (2001).

\bibitem{ross} S. Chakraborty, {\it Is Emergent Universe a Consequence of Particle Creation Process?}, Phys. Lett. {\bf B 732}, 81--84 (2014), arXiv:1403.5980 [gr-qc]; \newline
T. Harko and F.S.N. Lobo, {\it Irreversible Thermodynamic Description of Interacting Dark Energy -- Dark Matter Cosmological Models}, Phys. Rev. {\bf D 87}, 044018 (2013), arXiv:1210.3617 [gr-qc]; \newline
S.K. Biswas, W. Khyllep, J. Dutta, S. Chakraborty, {\it Dynamical Analysis of Interacting Dark Energy Model in the Framework of Particle Creation Mechanism}, Phys. Rev. {\bf D 95}, 103009 (2017), arXiv:1604.07636 [gr-qc]; \newline
S. Pan, J. de Haro, A. Paliathanasis, R.J. Slagter, {\it Evolution and Dynamics of a Matter Creation Model}, Mon. Not. Roy. Astron. Soc. {\bf 460(2)}, 1445-1456 (2016), arXiv:1601.03955 [gr-qc].

\bibitem{nie} R.I. Ivanov and E.M. Prodanov, {\it  Dynamical Analysis of an $n$--$H$--$T$ Cosmological Quintessence Real Gas Model with a General
Equation of State}, Int. J. Mod. Phys. {\bf A 33(03)}, 1850025 (2018).

\bibitem{viscosity} B.L. Hu, {\it Vacuum Viscosity Description of Quantum Processes in the Early Universe}, Phys. Lett. {A 90 (7)}, 375 (1982).


\bibitem{zimdahl2} W. Zimdahl, {\it Bulk Viscous Cosmology}, Phys. Rev. {\bf D 53(10)}, 5483 (1996), arXiv:astro-ph/9601189.

\bibitem{freaza} M.P. Freaza, R.S. de Souza, and I. Waga, {\it Cosmic Acceleration and Matter Creation}, Phys. Rev. {\bf D 66}, 103502 (2002).

\bibitem{br3} I. Brevik and G. Stokkan, {\it Viscosity and Matter Creation in the Early Universe}, Astrophys. Space Sci. {\bf 239}, 89--96 (1996).

\bibitem{maa} R. Maartens, {\it Causal Thermodynamics in Relativity (Lectures given at the Hanno Rund Workshop on Relativity and Thermodynamics,
University of Natal, June 1996)}, astro-ph/9609119.

\bibitem{lima} J.A.S. Lima, {\it Thermodynamics of Decaying Vacuum Cosmologies}, Phys.Rev. {\bf D 54}, 2571--2577 (1996), gr-qc/9605055.


\bibitem{gor}  A. Goriely, {\it Integrability and Non-integrability of Dynamical Systems}, World Scientific (2001).

\bibitem{car} S.M. Carroll, M. Hoffman,  and M. Trodden, {\it Can the dark energy equation-of-state parameter $\omega$ be less than $-1$?}, Phys. Rev.
{\bf D 68}, 023509 (2003), astro-ph/0301273.

\bibitem{aaa} J. Guckenheimer and P. Holmes, {\it Nonlinear Oscillations, Dynamical Systems, and Bifurcations of Vector Fields}, Appl. Math. Sci. {\bf 42}, Springer-Verlag Berlin New York (1986).

\end{thebibliography}
\end{document}